\documentclass[times,twocolumn,final,authoryear]{elsarticle}

\usepackage{jasr}
\usepackage{framed,multirow}

\usepackage{amssymb}
\usepackage{latexsym}

\usepackage{url}
\usepackage{xcolor}
\definecolor{newcolor}{rgb}{.8,.349,.1}

\usepackage[citebordercolor=white]{hyperref}

\usepackage{amsmath}
\usepackage{amsthm}
\usepackage{amssymb}
\usepackage{amsfonts}
\usepackage{bm}
\usepackage[overload]{empheq} 
\usepackage{fix-cm} 
\usepackage{lineno}
\usepackage{cleveref}
\usepackage[caption=false]{subfig}

\crefname{equation}{Eq.}{Eqs.}
\crefname{table}{Table}{Tables}
\crefname{figure}{Fig.}{Figs.}
\crefname{subsection}{Sec.}{Secs.}
\crefname{section}{Sec.}{Secs.}
\crefname{chapter}{Chap.}{Chaps.}
\crefname{appendix}{}{}
\newcommand{\norm}[1]{\lVert#1\rVert}
\renewcommand{\vec}[1]{{\text{\boldmath{$#1$}}}}
\newcommand{\virgolette}[1]{``#1''}
\newcommand*\arcsec{\ensuremath{^{\prime\prime}}}
\usepackage{makecell}
\usepackage{threeparttable}

\journal{Advances in Space Research}

\begin{document}

\verso{G. Littoriano \textit{et al.}}

\begin{frontmatter}

\title{Modelling of spacecraft apparent brightness \\ A study on OneWeb constellation satellites}

\author[1]{Gerardo \snm{Littoriano}}
\ead{gerardo.littoriano@mail.polimi.it}
\author[1]{Camilla \snm{Colombo}\corref{cor1}}
\cortext[cor1]{Corresponding author}
\ead{camilla.colombo@polimi.it}
\author[2]{Alessandro \snm{Nastasi}}
\ead{alessandro.nastasi@galhassin.it}
\author[3]{Carmelo \snm{Falco}}
\ead{carmelo.falco.galhassin@gmail.com}

\address[1]{Department of Aerospace Science and Technology (DAER), Politecnico di Milano, Via La Masa 34, Milano (MI), 20156, Italy}
\address[2]{GAL Hassin - Centro Internazionale per le Scienze Astronomiche, Via della Fontana Mitri, Isnello (PA), 90010, Italy\\ INAF – Osservatorio Astronomico di Palermo, piazza
del Parlamento 1, 90134, Palermo (Italy) 
}
\address[3]{GAL Hassin - Centro Internazionale per le Scienze Astronomiche, Via della Fontana Mitri, Isnello (PA), 90010, Italy\\ INAF - Osservatorio Astrofisico di Torino, via Osservatorio 20, 10025 Pino Torinese (Italy) 
}


\begin{abstract}
Artificial satellites orbiting around the Earth, under certain conditions, result to be visible even to the naked eye. The phenomenon of light pollution jeopardises the researching activities of the astronomical community: traces left by the objects are clear and evident and images for scientific purposes are damaged and deteriorated.

The development of a mathematical model able to estimate the satellite's brightness is required and it represents a first step to catch all the aspects of the reflection phenomenon. The brightness model (by Politecnico di Milano) will be exploited to implement a realistic simulation of the apparent magnitude evolution and it could be used to develop an archetype of new-generation spacecraft at low light-pollution impact.

Starting from classical photometry theory, which provides the expressions of radiant flux density of natural spherical bodies, the global laws describing flux densities and the associated apparent magnitude are exploited to generalise the analysis. The study is finally focused on three-dimensional objects of whatever shape which can be the best representation of the spacecraft geometry. To obtain representative results of the satellite brightness, a validation process has been carried on. The observation data of OneWeb satellites have been collected by GAL Hassin astronomical observatory, settled in Isnello, near Palermo. The observations were carried out in order to map the satellites brightness at various illumination conditions, also targeting a single satellite across its different positions on the sky (i.e., during its rise, culmination and setting).


\end{abstract}


\end{frontmatter}




\section{Introduction}
\label{Intro}
\begin{figure}[htp]
\resizebox{\hsize}{!}{
\subfloat[]{\includegraphics[width = 0.5\textwidth]{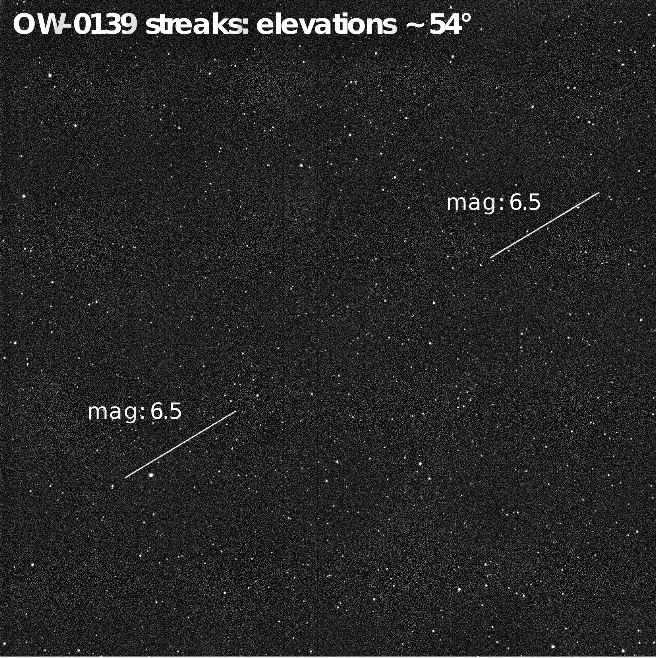}}\quad
\subfloat[]{\includegraphics[width = 0.5\textwidth]{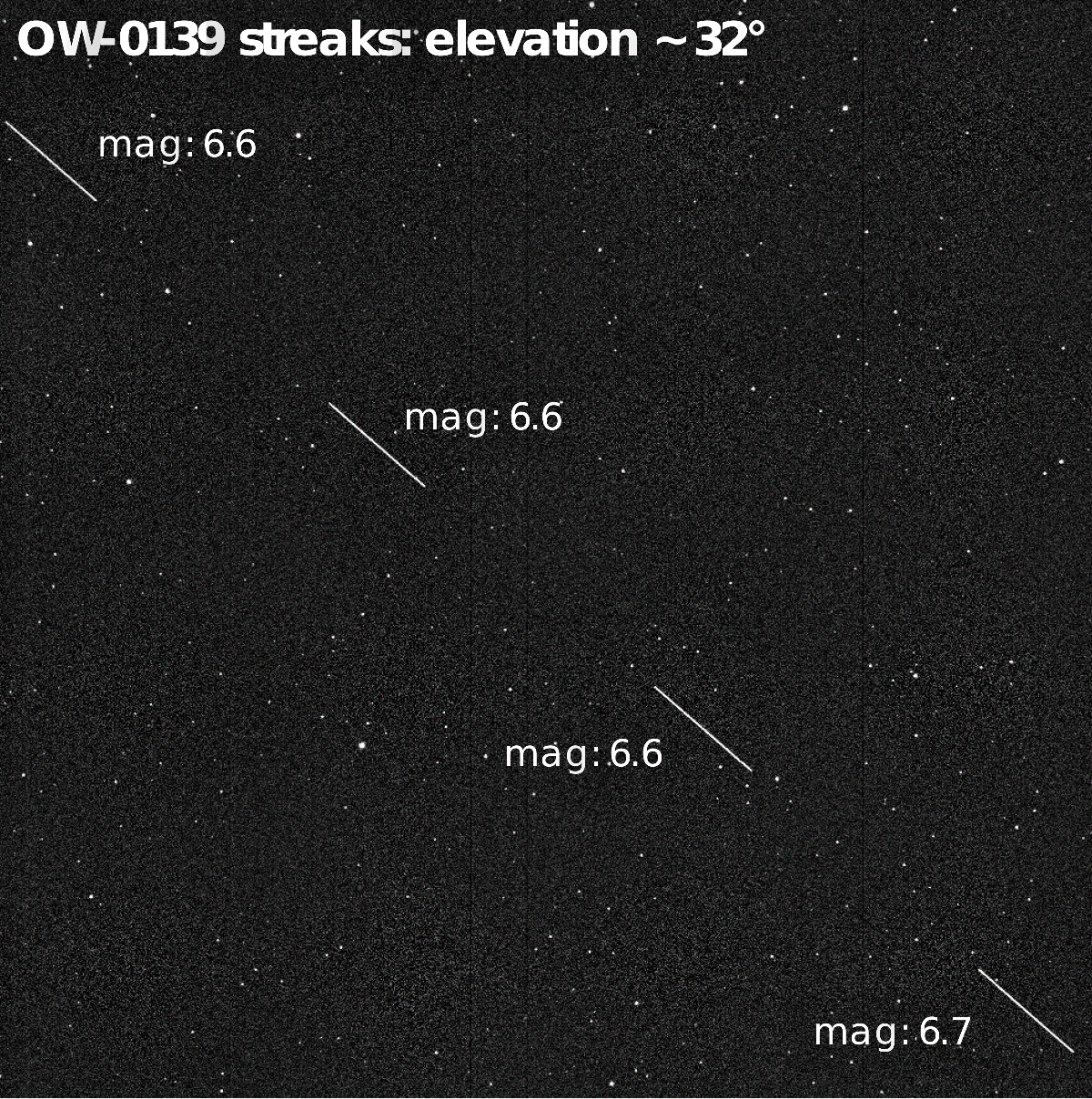}}\quad}
\caption{Trails of OneWeb satellites. Example of observations made by GAL Hassin astronomical observatory.}
\label{fig: C1 - OW}
\end{figure}

The brightness of artificial objects orbiting around the Earth affects the visibility from ground of the celestial bodies, such as stars or planets, making harder the observations and studies performed by astronomers and scientists. The launch of hundreds of satellites represents a serious problem from the optical astronomy point of view. Since the deployment of the first Starlink satellites in 2019, the possibility to spot the flying units to the naked eye is become very high also in light polluted cities \citep{space_debris_conf}. 
In particular conditions, some satellites are able to reflect the sunlight and to be figured as trails in astronomical observations. Furthermore, Low-Earth-Orbit (\emph{LEO}) satellites appear brighter than other spacecraft due to the smallest distances from the observers.
Satellite trails cannot easily removed from observation images and, therefore, the astronomer work is jeopardised by these effects. The solution to this issue must be identified and adopted for the next generation satellites in order to preserve and enhance the precious contribution from the astronomical community.\\
The first step of this analysis is represented by a brightness model, developed by Politecnico di Milano, which takes into account object material's reflecivities, orbital motion of the satellite, its geometrical areas and, finally, its attitude (angles between sources/observer and surfaces), giving as a result the spacecraft apparent magnitude.\\
GAL Hassin astronomical observatory ($\lambda = 14^{\circ} 01' 14.2''$, $\phi = 37^{\circ} 56' 21.8''$) will play a crucial role for the proposed study. It is an observatory located in Isnello, near Palermo (Italy). The peculiarities of GAL Hassin observatory can be found into the low-polluted sky, in terms of light sources, surrounding the park, favorable weather conditions that allow easier and frequent observations, and low-latitude location, highlighting the accessibility to observe the Milky Way center. \\
GAL Hassin observatory provides emblematic observations of OneWeb satellites in order to compare them with the results of the developed brightness model with the aim of validating it. 
An example of observation data (already subjected to photometric analysis) exploited for the validation test is pointed out by \cref{fig: C1 - OW}. In these images, the trails left by the satellite passage negatively affect the astronomical observations of surrounding stars and celestial objects.\\
\\
Once the model has been built up and validated, the research of causes and sources of possible luminous flares by artificial satellites can be set up and the next-generation satellites at low light-pollution impact can be designed.\\

\subsection{Literary review}
To understand and analyse the brightness of artificial and generally-shaped bodies, it is useful to review theories and models adapted to spherical celestial objects. \citet{Planetary_photometry} derived the theory of planetary photometry and the relationships between the apparent magnitude of a planet and the phase angle (Sun - planet - observer angle) together with the properties of the celestial body's surface. Indeed, first of all, the physical quantities that are involved in photometry are introduced: luminous power/flux, luminous intensity, illuminance, flux density and luminance. The definition of Bidirectional Reflectance-Distribution Function (\emph{BRDF}) is presented and its correlation with the aforementioned entities. The reflectance is figured out as well (concepts discussed two years before by \citet{Nicodemus}). \\
\citet{Planetary_photometry} provides particular laws to derive simple analytical expressions of the \emph{BRDF}, exploited for the aim of this work:

\begin{itemize}
\item \textbf{Lambert's Law}. The radiance of an emitting surface is independent on the observation angle while the angle of incidence affects the radiance of the reflecting surface. Under this hypothesis, it is possible derive the statement of constant \emph{BRDF}.
\item \textbf{Lommel-Seeliger Law}. In this model, it is assumed that the penetrated light through the surface of an object is exponentially absorbed and then it is scattered in any direction, emerging by the surface as well. Here, the \emph{BRDF} is inversely proportional to the summation of the incidence and viewing angle's cosines.
\item \textbf{Area Law}. The \emph{BRDF} is assumed to be proportional to the secant of the angle of incidence. However there is a proof demonstrating the impossibility of existence of a real surface following the area law for all incidence angles. 
\end{itemize}

Finally, in the theory of planetary photometry, the analysis is specialised for spheres. In details, the phase angle is introduced and for each model the phase law (function of phase angle exclusively) is described.

\citet{Roccioletti} exploited the model above to describe an artificial balloon's brightness. Both specular and diffusive reflections have been studied, together with the brightness associated with the earthshine contribution. A sensitivity analysis is performed to understand which are the most influencing orbital and seasonal parameters. Indeed, the visibility of the spacecraft from nine different cities and the visibility time slots should be maximised in order to accomplish the mission goals, by optimising the design of the satellite orbital motion which, in the end, results to be a Sun-synchronous orbit. Within Roccioletti's dissertation, the expression of the apparent magnitude with respect to the Sun is presented by \citet{apparent_magnitude}.

\citet{mallama_panel} suggests a simple model for a flat panel orbiting around the Earth. It is a very simple schematisation of a Starlink satellite, being characterised by a quasi-nadir-pointing flat panel, equipped with a single solar array. In that paper, the relative magnitude of the satellites depends only on the zenith angle and the orientation of the panel with respect to the Sun. Indeed, for non-spherical objects, the attitude of the body strongly affects the reflectivity of the object itself. Finally, the sensitivity to the Sun declination and elevation with respect to the local horizon is analysed.

A more realistic and more detailed model for Starlink's satellites is provided by \citet{Cole}. Four different contribution to the satellite brightness are analysed: diffusive reflection from the frontal face of the solar array;
diffusive reflection from the back face of the solar array (even if this surface is not directly illuminated by the Sun, a small percentage of the incident light upon the frontal one passes inside the small aperture between solar cells, being reflected in diffusive way);
diffusive reflection
from the spacecraft base;
specular reflection from the satellite base.

All these inputs are linked with peculiar angles between Sun, the surface's normal vectors and observer positions and a semi-empirical formula is retrieved for the reflected flux. Some coefficients directly associated with the surface's areas and with the corresponding reflectivities are optimised in order to fit the observation data. 

New well-defined model not involving any fitting process is presented by \citet{New_model}, in which a very detailed radiant flux density and the associated visual magnitude are defined, depending only on geometrical, physical and orbital features. Indeed, Starlink satellites, modelled as flat nadir-pointing surfaces equipped with a single solar array perpendicular to the former, are analysed and the consequent radiant flux density and associated apparent magnitude are function of material characteristics (reflecitivity), panel areas and \emph{BRDF} which depends on the angles between the surface normal vectors and the Sun/observer locations upon the globe. Unfortunately, only two dimensions of the satellite are taken into account and this approximation is good enough only for real flat spacecraft, as the Starlink units appear.


\subsection{Contributions}
The work carried out in this article is the offspring of a collaboration between Politecnico di Milano, GAL Hassin astronomical observatory and OneWeb company. The necessity to understand the causes of satellite luminous peaks and to develop new technologies able to make the new generations of spacecraft fainter leads to the creation of the collaboration. 
Once the brightness evolution is analysed and the main parameters affecting its variations are identified, causes of flares and luminous peaks could be investigated. The technologies to be adopted and implemented in the next generation of spacecraft should conform the results coming out from this analysis. 
Therefore, the first step is the development of a tool able to estimate the brightness of artificial objects.


\section{Orbital dynamics and brightness models}
\label{Models}
\subsection{Orbital dynamics}

The orbital motion of each satellite is retrieved by propagating the Two Line Elements (TLEs). SGP4 is the propagation model exploited to estimate spacecraft position in ECI reference frame. 

\subsection{Photometric model}

First of all, the brightness of an object is measured by the apparent magnitude ($m$), as shown by \citet{apparent_magnitude}:

\begin{equation}\label{eq: Magnitude}
    m = -2.5\log_{10}\biggl(\dfrac{F}{F_0}\biggr)
\end{equation}

\noindent where $F_0$ is the incident luminous flux upon the object while $F$ is the reflected one.
Furthermore, it is possible to select a specific body and its flux as reference. Usually, Sun is chosen and, in this case, the apparent magnitude is given by (\citet{apparent_magnitude}):

\begin{equation}\label{eq: magnitude Sun}
   m = m_{\odot}-2.5\log_{10}\biggl(\dfrac{F}{F_{\odot}}\biggr) \hspace{0.3cm} \text{in which} \hspace{0.3cm} m_{\odot} = -26.74
\end{equation}
It is worth to recall that the lower $m$, the brighter is the perception of the object. 

\begin{figure}[htp]
    \centering
    \includegraphics[width = 0.3\textwidth]{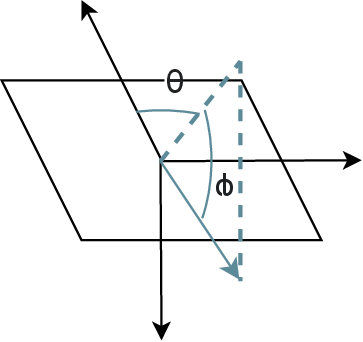}
    \caption{Definition of azimuth and elevation angles.}
    \label{fig: Spherical coordinates}
\end{figure}

There exist several models to study the satellite magnitudes, used for observations. A general model from fundamental definitions is retrieved by \citet{Planetary_photometry}.\\
The planetary photometry is a science which analyses the intensity of light perceived by human eyes coming from planetary objects, such as Solar system's planets or asteroids. By following the same steps shown by \citet{Planetary_photometry}, a generic direction can be expressed by means of two angles: the azimuth ($\theta$) and the elevation ($\phi$) angles, as shown in \cref{fig: Spherical coordinates}. 

Considering the irradiance $E(\theta_i,\phi_i)$ of a surface due to an incident radiation coming from the direction ($\theta_i$,$\phi_i$) and the radiance $L(\theta_r,\phi_r$) in the direction of reflection ($\theta_r$,$\phi_r$), the Bidirectional Reflectance Distribution Function or \emph{BRDF}, $f_r(\theta_i,\phi_i;\theta_r,\phi_r)$ is given by \cref{eq: BRDF}:

\begin{equation}\label{eq: BRDF}
    f_r(\theta_i,\phi_i;\theta_r,\phi_r) = \dfrac{L(\theta_r,\phi_r)}{E(\theta_i,\phi_i)}
\end{equation}

The bidirectional hemispherical reflectance $\rho(\theta_i,\phi_i)$ is defined as the ratio between the reflected power per unit area at a point on a surface and the incident radiance. In other words, it is shown in \cref{eq: rho}.

\begin{equation}\label{eq: rho}
    \rho(\theta_i,\phi_i) = \int_{\theta=0}^{2\pi}\int_{\phi=0}^{\frac{\pi}{2}}f_r(\theta_i,\phi_i;\theta_r,\phi_r)\cos(\theta_r)\sin(\theta_r)d\theta_rd\phi_r
\end{equation}

Now, some helpful models shall be introduced in order to simplify the expression of the \emph{BRDF} and to derive analytic formulations for luminous fluxes.

\begin{figure}
    \centering
    \includegraphics[width = 0.3\textwidth]{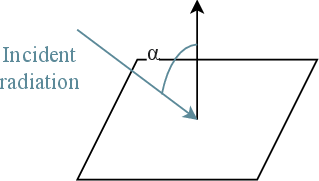}
    \caption{Incident radiation and angle between surface normal vector.}
    \label{fig: alpha}
\end{figure}

\paragraph{Lambert's law} A Lambertian emitting surface is characterised by a radiance independent on the angle of observation. This means that this principle is valid for each angle of incidence. By taking $\alpha$ as the angle between the normal of a Lambertian element area and the incoming radiation (see \cref{fig: alpha}), the radiant intensity can be expressed as \cref{eq: Lambert intensity}.

\begin{equation}\label{eq: Lambert intensity}
    I(\alpha) = I(0)\cos(\alpha)
\end{equation}

$I(0)$ is the radiant intensity associated with normal incident radiation and, integrating on an hemisphere, the power irradiated over its external surface by a Lambertian emitting surface is given by \cref{eq: Lambert power}.

\begin{equation}\label{eq: Lambert power}
    P = \pi I(0)
\end{equation}

For this kind of surfaces, the \emph{BRDF} is a constant, independent on the directions of the incident and reflected radiations:

\begin{equation}\label{eq: Lambert fr}
    f_r = \gamma
\end{equation}

\noindent It follows from \cref{eq: Lambert fr,eq: rho} that the \emph{BRDF} of a Lambertian surface is given by \citet{Planetary_photometry}:

\begin{equation}\label{eq: Lambert BRDF}
    f_r = \frac{\rho}{\pi}
\end{equation}

\paragraph{Lommel-Seeliger's Law} The Lommel-Seeliger's law suggests that the light penetrates through the object's surface and it is exponentially absorbed. Each volume element, then, scatters the radiation in all the directions and part of this latter emerges from the surface. Under this hypothesis, the \emph{BRDF} is supposed to be expressed as shown in \cref{eq: LS fr}, in which $f(\theta_i,\theta_r)$ is a function which depends only on the azimuth angles of incidence and reflection but not on elevations. $\gamma$, instead, is a constant having a dimension of $sr^{-1}$.

\begin{equation}\label{eq: LS fr}
    f_r(\theta_i,\phi_i;\theta_r,\phi_r) = \dfrac{\gamma f(\theta_i,\theta_r)}{\cos(\theta_i)\cos(\theta_r)}
\end{equation}

The function $f(\theta_i,\theta_r)$ and the resulting \emph{BRDF} for a Lommel-Seeliger surface are shown in \cref{eq: LS BRDF}.

\begin{equation}\label{eq: LS BRDF}
    \begin{cases}
    \medskip
    f(\theta_i,\theta_r) = \dfrac{\cos(\theta_i)\cos(\theta_r)}{\cos(\theta_i)+\cos(\theta_r)}\\
    f_r(\theta_i,\phi_i;\theta_r,\phi_r)=\dfrac{\gamma}{\cos(\theta_i)+\cos(\theta_r)}\\
    \end{cases}
\end{equation}

Finally, the directional hemispherical reflectance, within the Lommel-Seeliger's law assumption, can be expressed as reported in \cref{eq: LS rho}.

\begin{equation}\label{eq: LS rho}
    \rho(\theta_i,\phi_i) = 2\pi\gamma\{1-\cos(\theta_i)\ln[1+\sec(\theta_i)]\}
\end{equation}

Note that $\rho(\theta_i,\phi_i)$ monotonically increases from $2\pi\gamma(1-\ln(2))$ at normal incidence, and $2\pi\gamma$ at grazing incidence. Since $\rho(\theta_i,\phi_i)$, by conservation of energy, cannot be larger than the unity, $\gamma$ must be lower or at least equal than $1/(2\pi)$.

\paragraph{Area law} The area law makes the following hypothesis:

\begin{equation}\label{eq: AL fr}
    f_r(\theta_i,\phi_i;\theta_r,\phi_r)=\gamma \sec(\theta_i)
\end{equation}

This means that the directional hemispherical reflectance is given by \citet{Planetary_photometry}:

\begin{equation}\label{eq: AL rho}
    \rho(\theta_i,\phi_i) = \pi\gamma\sec(\theta_i)
\end{equation}

As already stated for the Lommel-Seeliger assumption, the conservation of energy suggests that the directional hemispherical reflectance cannot exceed the unity. Therefore, the area law is not applicable for real surfaces. Nevertheless, focusing on spheres, the area law is of interest whenever $\gamma = 1/(2\pi)$.

From the definition of \emph{BRDF} (\cref{eq: BRDF}) and knowing that the irradiance $E(\theta_i,\phi_i)$ due to an incident radiation is associated with the incoming radiant flux density $F_0$ through the following relationship $E(\theta_i,\phi_i) = F_0\cos(\theta_i)$, the radiance of a generic point in the direction of the observer $L(\theta_r,\phi_r)$ is expressed as shown in \cref{eq: L_sphere}.

\begin{equation}\label{eq: L_sphere}
\begin{split}
    L(\theta_r,\phi_r) = f(\theta_i,\phi_i;\theta_r,\phi_r)E(\theta_i,\phi_i) = \\
    = F_0f(\theta_i,\phi_i;\theta_r,\phi_r)\cos(\theta_i)
\end{split}
\end{equation}

By taking an elementary area $dA$ and its projection onto the direction of the observer, $\cos(\theta_r)dA$, the elementary radiant intensity ($dI$) is expressed as \cref{eq: I}. The associated radiant flux density $dF$ is simply retrieved by dividing this quantity by the square of the distance between object and observer $r_o$ (\cref{eq: F}).

\begin{equation}\label{eq: I}
\begin{split}
    dI(\theta_i,\phi_i;\theta_r,\phi_r) = L(\theta_r,\phi_r)\cos(\theta_r)dA = \\ = F_0f(\theta_i,\phi_i;\theta_r,\phi_r)\cos(\theta_i)\cos(\theta_r)dA 
\end{split}
\end{equation}
\begin{equation}\label{eq: F}
\begin{split}
    dF(\theta_i,\phi_i;\theta_r,\phi_r) = \frac{dI(\theta_i,\phi_i;\theta_r,\phi_r)}{r_o^2} = \\ = \frac{F_0}{r_o^2}f(\theta_i,\phi_i;\theta_r,\phi_r)\cos(\theta_i)\cos(\theta_r)dA
\end{split}
\end{equation}


\subsection{Planar models of brightness}

Focusing on a plane, the attitude of this latter plays a fundamental role in the reflection of the incident radiation and photometric quantities are function of the incident and viewing angles ($\theta_i,\theta_r)$. The integration over the surface of the object is quite trivial, being the normal vector to the surface pointing towards a constant direction. For a plane, the angles between the surface's normal vector and the sub-source and sub-observer directions are constant and may be taken out from the integration process. 

\paragraph{Lambert's plane} For a Lambertian reflector, the \emph{BRDF} is a constant value and the respective radiant intensity can be expressed as shown in \cref{eq: I Lambert plane}.

\begin{equation}\label{eq: I Lambert plane}
\begin{split}
    I(\theta_i,\theta_r) = F_0\dfrac{\rho}{\pi}\cos(\theta_i)\cos(\theta_r)\int dA
    = \\ = F_0\dfrac{\rho A}{\pi}\cos(\theta_i)\cos(\theta_r)
\end{split}
\end{equation}

\noindent The consequent radiant flux density and its related visual magnitude can be computed as represented by \cref{eq: Magnitude Lambert plane}.

\begin{equation}\label{eq: Magnitude Lambert plane}
    \begin{cases}
    \smallskip
    F(\theta_i,\theta_r) = \dfrac{I(\theta_i,\theta_r)}{r_o^2}=F_0\dfrac{\rho A}{\pi r_o^2}\cos(\theta_i)\cos(\theta_r)\\
    m = -26.74-2.5\log_{10}\biggl[\dfrac{\rho A}{\pi r_o^2}\cos(\theta_i)\cos(\theta_r)\biggr]\\
    \end{cases}
\end{equation}

\paragraph{Lommel-Seeliger's plane} The radiant intensity for a Lommel-Seeliger planar reflector is given by \cref{eq: I LS plane}.

\begin{equation}\label{eq: I LS plane}
\begin{split}
    I(\theta_i,\theta_r) = F_0\gamma\dfrac{\cos(\theta_i)\cos(\theta_r)}{\cos(\theta_i)+\cos(\theta_r)}\int dA = \\ = F_0\gamma A\dfrac{\cos(\theta_i)\cos(\theta_r)}{\cos(\theta_i)+\cos(\theta_r)}
\end{split}
\end{equation}

At this point, the radiant flux density and magnitude associated with the Lommel-Seeliger's law are provided by \cref{eq: Magnitude LS plane}.

\begin{equation}\label{eq: Magnitude LS plane}
    \begin{cases}
    \smallskip
    F(\theta_i,\theta_r) = \dfrac{I(\theta_i,\theta_r)}{r_o^2}=F_0\dfrac{\gamma A}{r_o^2}\dfrac{\cos(\theta_i)\cos(\theta_r)}{\cos(\theta_i)+\cos(\theta_r)}\\
    m = -26.74-2.5\log_{10}\biggl[\dfrac{\gamma A}{r_o^2}\dfrac{\cos(\theta_i)\cos(\theta_r)}{\cos(\theta_i)+\cos(\theta_r)}\biggr]\\
    \end{cases}
\end{equation}

\paragraph{Area law plane} Lastly, looking at the area law, the radiant intensity at $(\theta_i,\theta_r)$ is given by \cref{eq: I AL plane}.

\begin{equation}\label{eq: I AL plane}
    I(\theta_i,\theta_r) = F_0\gamma\cos(\theta_r)\int dA = F_0\gamma A\cos(\theta_r)
\end{equation}

It is important to highlight that, in this case, the percentage of perceived light does not depend on the angle of incidence but on the position of the observer only. The area law's flux density and magnitude for a planar surface are written in \cref{eq: Magnitude AL plane}.

\begin{equation}\label{eq: Magnitude AL plane}
    \begin{cases}
    \smallskip
    F(\theta_i,\theta_r) = \dfrac{I(\theta_i,\theta_r)}{r_o^2}=F_0\dfrac{\gamma A}{ r_o^2}\cos(\theta_r)\\
    m = -26.74-2.5\log_{10}\biggl[\dfrac{\gamma A}{ r_o^2}\cos(\theta_r)\biggr]\\
    \end{cases}
\end{equation}

\begin{figure}[htp]
    \centering
    \subfloat[][Lambert's law\label{fig: Lambert plane}]{\includegraphics[width = 0.4\textwidth]{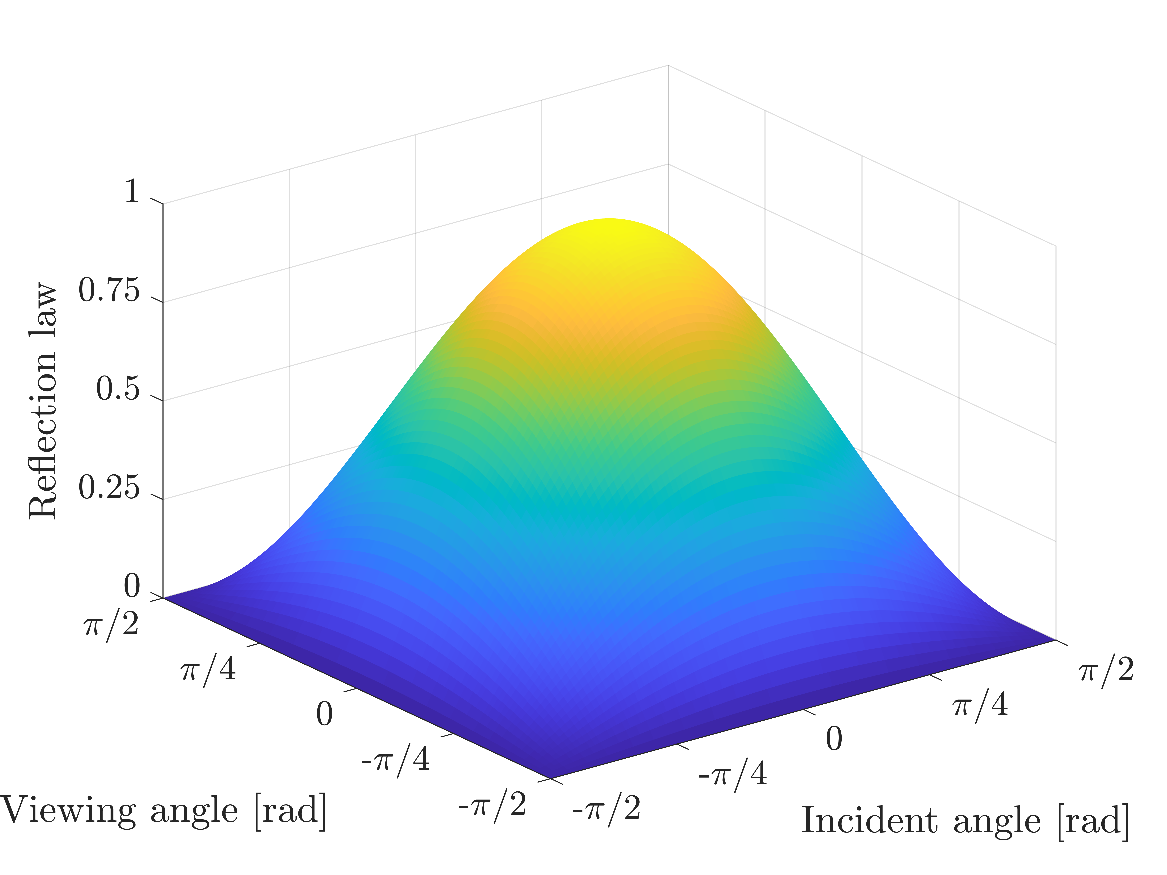}}\quad
    \subfloat[][Lommel-Seeliger's law\label{fig: LS plane}]{\includegraphics[width = 0.4\textwidth]{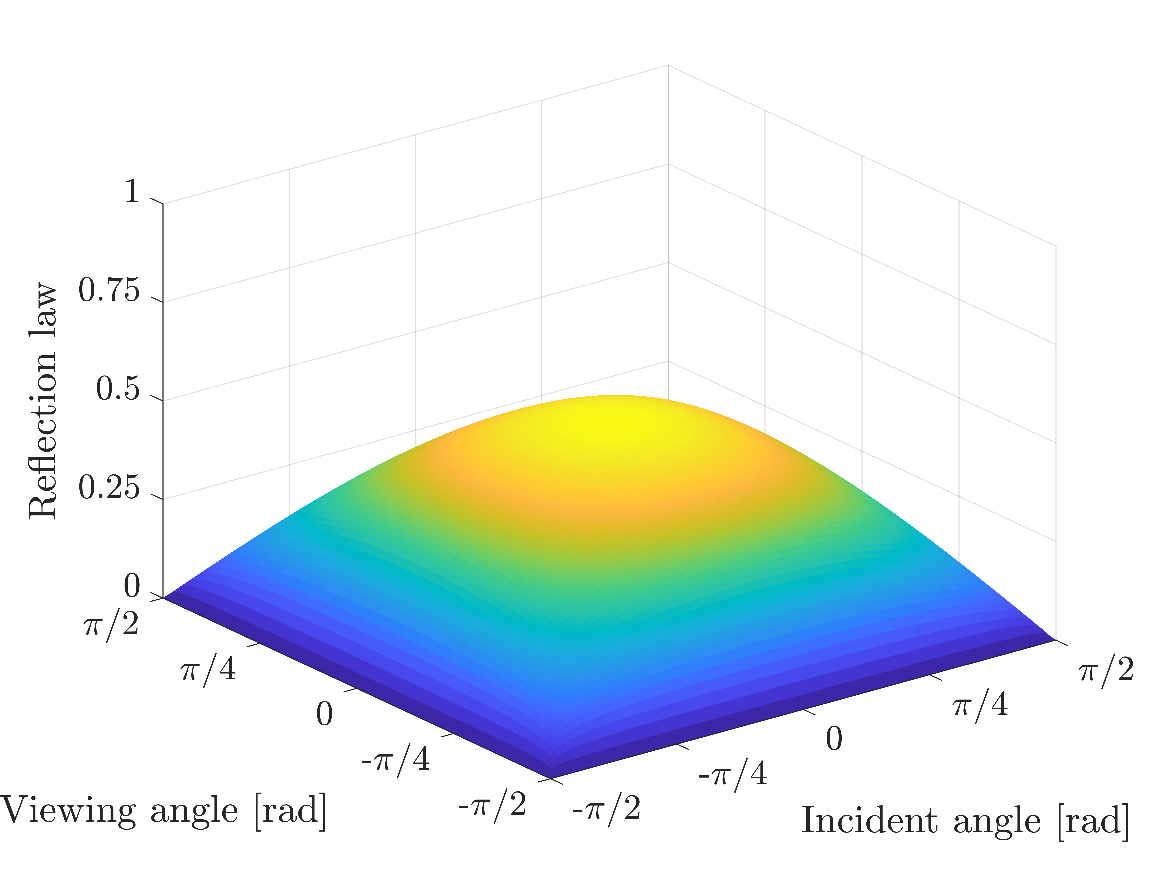}}\quad
    \subfloat[][Area law\label{fig: AL plane}]{\includegraphics[width = 0.4\textwidth]{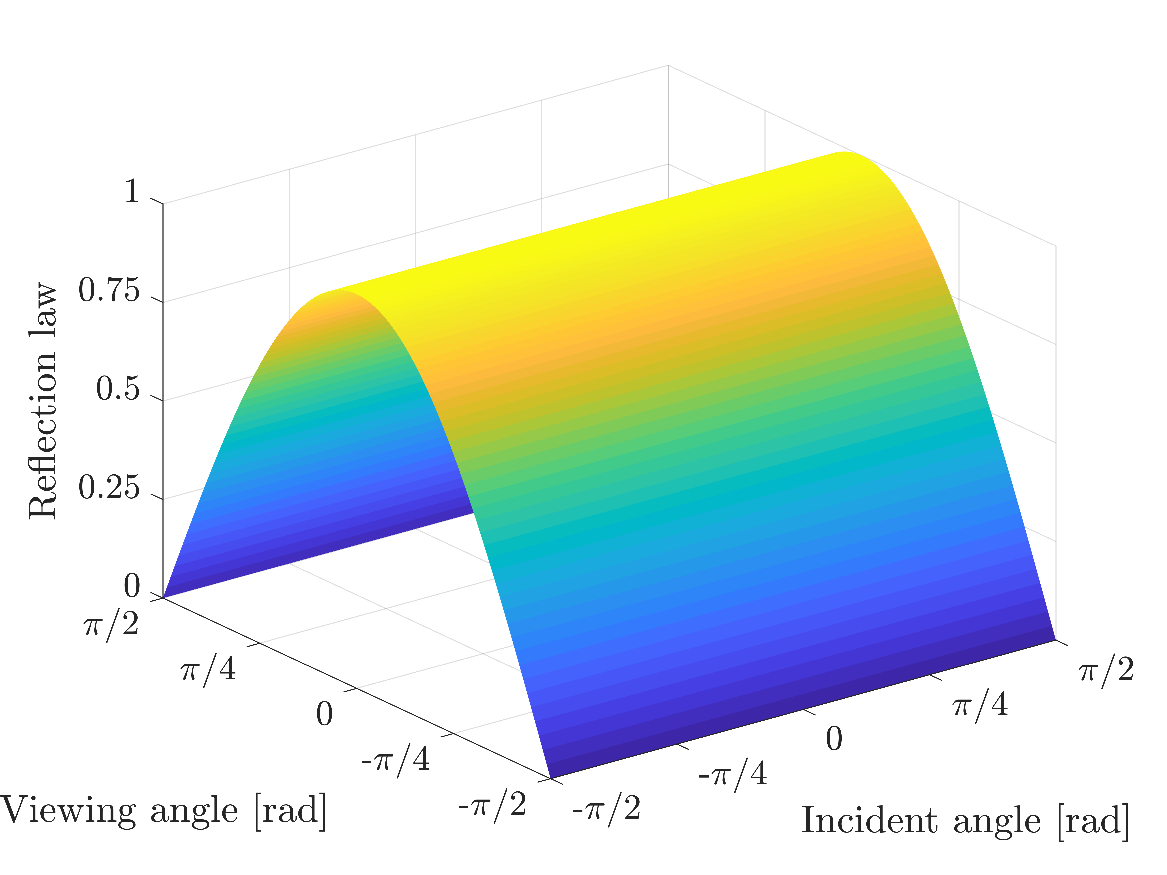}}\quad
    \caption{Reflection laws of planar surfaces.}
    \label{fig: Plane}
\end{figure}

To summarise the reflection of a planar surface which is modelled by following the three law explained previously, \cref{fig: Plane} is built up. In \cref{fig: LS plane} it is clear that the reflection fraction never reaches the unity, neither when the radiation source or the observer are perfectly aligned with the plane's normal vector but it equalises 50$\%$. The area law, depicted in \cref{fig: AL plane}, shows the non-dependence on the incident angle. The most complete law seems to be the Lambert's one, portrayed in \cref{fig: Lambert plane}, which is selected to model the brightness of the under analysis spacecraft.

Past works undertaken in this direction (brightness analysis for planar surfaces), described by \citet{New_model}, exploit the Lambert's law to estimate the apparent magnitude of flat surfaces.

\subsection{2D planar surface}\label{subsec: 2D plane}
Here the Lambert law is applied to a two dimensional plane that is reflecting the sunlight and the earthshine. As an example, for a nadir-pointing plane the normal vector to the surface, $\vec{\hat{n}}$, is opposite to the radial vector \vec{r} which links the satellite with the Earth-Centered-Intertial (\emph{ECI}) reference frame's center: $\vec{\hat{n}} = -\vec{r}/\norm{\vec{r}}$.

Knowing the position of the Sun with respect to the Earth \vec{r_{\odot}} (knowing its right ascension, $RA_{\odot}$, and declination, $Dec_{\odot}$), it is possible to reconstruct the Sun's location as against the satellite (\vec{{r_{SC\rightarrow\odot}}}) as

\begin{equation}\label{eq: r_sun}
    \vec{r_{\odot}} = a_{\odot}
    \begin{Bmatrix}
    \cos(RA_{\odot})\cos(Dec_{\odot})\\
    \sin(RA_{\odot})\cos(Dec_{\odot})\\
    \sin(Dec_{\odot})\\
    \end{Bmatrix} \rightarrow
    \vec{r_{SC\rightarrow \odot}} = \vec{r_{\odot}}-\vec{r}
\end{equation}
\noindent where $a_{\odot}$ is the semi-major axis of the orbit that the Earth traces around the star.

The incidence and the viewing angles are respectively the angles between the plane's normal direction and the vectors that define the source and observer locations. Looking at the radiation source, i.e. the Sun, the incident angle is given by \cref{eq: theta_i}

\begin{equation}\label{eq: theta_i}
    \cos(\theta_i) = \dfrac{\vec{r_{S/C\rightarrow \odot}}\cdot\vec{\hat{n}}}{\norm{\vec{r_{S/C\rightarrow \odot}}}}
\end{equation}

Focusing now on the observer, knowing its location in terms of latitude, $\phi_c$, and longitude, $\lambda_c$, on the Earth, the site position vector $\vec{r_c}$ can be written as

\begin{equation}\label{eq: r_c}
    \vec{r_c} = R_{\oplus}
    \begin{Bmatrix}
    \cos(\lambda_c+\omega_{\oplus}t)\cos(\phi_c)\\
    \sin(\lambda_c+\omega_{\oplus}t)\sin(\phi_c)\\
    \sin(\phi_c)\\
    \end{Bmatrix}
\end{equation}
where $\omega_{\oplus}$ is the mean angular velocity of the Earth and $R_\oplus$ is its mean radius.

The relative position vector of the satellite with respect to the observation site, \vec{r_o}, is expressed in \cref{eq: r_o}. Therefore, the viewing angle for the estimation of the radiant flux density is shown in \cref{eq: theta_r}.

\begin{equation}\label{eq: r_o}
    \vec{r_o} = \vec{r}-\vec{r_c}
\end{equation}
\begin{equation}\label{eq: theta_r}
    \cos(\theta_r) = -\dfrac{\vec{r_o}\cdot\vec{\hat{n}}}{\norm{\vec{r_o}}}
\end{equation}

\begin{figure}[htp]
    \centering
    \includegraphics[width=0.4\textwidth]{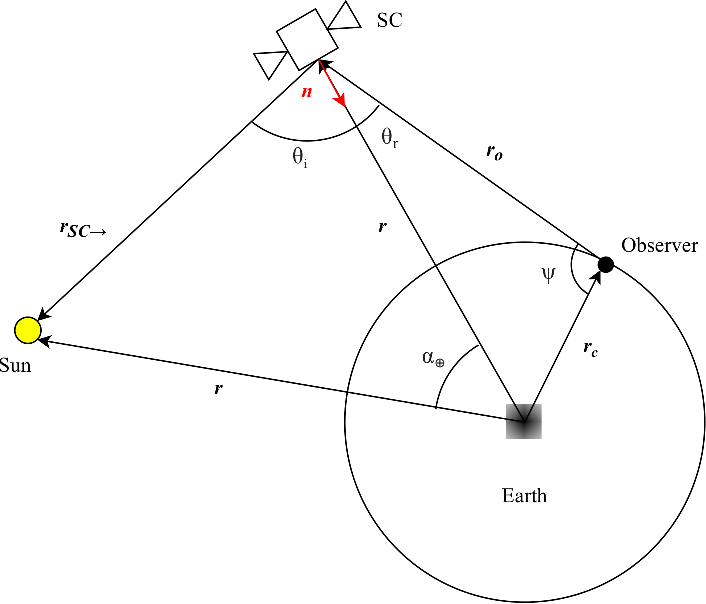}
    \caption{Satellite-Earth center-Observation site system.}
    \label{fig: EC-SC-GH}
\end{figure}

For the record, the angle between the observation site position vector and the one representing the relative satellite location compared to that is directly associated with the elevation of the satellite in Topocentric Horizon or South-East-Zenith (\emph{SEZ}) reference frame. By introducing an angle $\psi$ such that 
\begin{equation}\label{eq: psi}
    \psi = \pi - \dfrac{\vec{r_c}\cdot\vec{r_o}}{\norm{\vec{r_c}}\norm{\vec{r_o}}}
\end{equation}
\noindent then $\delta = \psi-\pi/2$ is retrieved. In \cref{fig: EC-SC-GH} a schematic view of the whole system is depicted in order to facilitate the comprehension of the scenario.

The radiant flux density associated with the Sun contribution that results from these quantities is given by \cref{eq: F 2D Sun}. In this equation, $\rho$ is the panel reflectance, $A$ is the surface area while $F_{\odot}$ represents the radiant flux density coming from the star.

\begin{equation}\label{eq: F 2D Sun}
    F_{S/C,\odot} = F_{\odot}\dfrac{\rho A}{\pi \norm{\vec{r_o}}^2}\cos(\theta_i)\cos(\theta_r)
\end{equation}

Furthermore, for the earthshine contribution, the reflection of the planet is modelled as shown by \citet{Earthshine}. In addition, the radiant flux density that results from the Earth flux is shown by \cref{eq: F 2D Earth}. In this case, the incidence angle, $\theta_{i,\oplus}$ is represented by the angular distance between the Earth position with respect to the satellite and the normal direction to the panel. In the nadir-pointing panel case, being this latter's normal direction parallel to its position vector, the incidence angle is null.

\begin{equation}\label{eq: F 2D Earth}
    \begin{cases}
    \smallskip
    \psi(\alpha_{\oplus}) =  [\sin(\alpha_{\oplus}+(\pi-\alpha_{\oplus})\cos(\alpha_{\oplus})]\\
    F_{\oplus} = \dfrac{2}{3}A_{\oplus}\dfrac{R_{\oplus}^2F_{\odot}}{\pi \norm{\vec{r_o}}^2}\psi(\alpha_{\oplus})\\
    F_{S/C,\oplus} = F_{\oplus}\dfrac{\rho A}{\pi \norm{\vec{r_o}}^2}\cos(\theta_{i,\oplus})\cos(\theta_r)\\
    \end{cases}
\end{equation}

Finally, the brightness of the nadir-pointing object is measured by the visual magnitude computed as shown in \cref{eq: 2D magnitude}. 

\begin{equation}\label{eq: 2D magnitude}
    m = -26.74-2.5\log_{10}\biggl(\dfrac{F_{S/C,\odot}+F_{S/C,\oplus}}{F_{\odot}}\biggr)
\end{equation}


\subsection{3D prism with a nadir-pointing face}
Looking at three dimensional objects, more specifically a prisms, six faces can be identified, each of them characterised by a normal vector to the surface. All the steps described in \cref{subsec: 2D plane} can be applied to each surface in which the proper normal vector is taken into account. 

\begin{figure}[htp]
\centering
    \includegraphics[width = 0.4\textwidth]{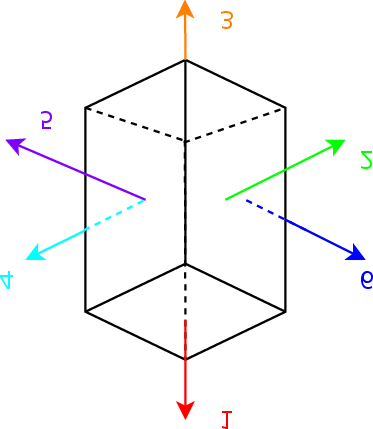}
    \caption{3D prism: Identification of surfaces and normal vectors.}
    \label{fig: prism}
\end{figure}

Initially, it is necessary to define a convention for the identification of the prism faces. In \cref{fig: prism}, it is possible to recognise the normal vectors to the surfaces in which:
\begin{enumerate}
    \item The first surface is defined to be the nadir-pointing plane whose normal direction results to be opposite to the position vector of the satellite:
    \begin{equation}\label{eq: n1}
        \vec{\hat{n}_1} = -\dfrac{\vec{r}}{\norm{\vec{r}}}
    \end{equation}
    \item The second surface is identified to be the one whose normal vector is aligned with the tangential direction of the orbital motion. If the ellipse's eccentricity was null, it would be the direction of the spacecraft velocity, $\vec{v}$. Particularly, the specific angular momentum direction, $\vec{\hat{h}}$, is given by:
    \begin{equation}\label{eq: h}
        \vec{\hat{h}} = \dfrac{\vec{r}\times\vec{v}}{\norm{\vec{r}\times\vec{v}}}
    \end{equation}
    The tangential direction, $\vec{\hat{t}}$ is then retrieved as shown in \cref{eq: t}.
    \begin{equation}\label{eq: t}
        \vec{\hat{t}} = \dfrac{\vec{\hat{h}}\times\vec{r}}{\norm{\vec{\hat{h}}\times\vec{r}}}
    \end{equation}
    Therefore, the normal vector to the second surface is expressed as
    \begin{equation}\label{eq: n2}
        \vec{\hat{n}_2} = \vec{\hat{t}}
    \end{equation}
    \item The third normal direction is opposite to the first one. Consequently it is directed as the position vector of the satellite.
    \begin{equation}\label{eq: n3}
        \vec{\hat{n}_3} = -\vec{\hat{n}_1}
    \end{equation}
    \item The fourth normal vector is opposite to the tangential direction and so to the second normal vector:
    \begin{equation}\label{eq: n4}
        \vec{\hat{n}_4} = -\vec{\hat{n}_2}
    \end{equation}
    \item The fifth normal vector is directed as the specific angular momentum direction.
    \begin{equation}\label{eq: n5}
        \vec{\hat{n}_5} = \vec{\hat{h}} 
    \end{equation}
    \item Finally, the sixth surface's normal vector points in the opposite direction of $\vec{\hat{n}_5}$:
    \begin{equation}\label{eq: n6}
        \vec{\hat{n}_6} = -\vec{\hat{n}_5}
    \end{equation}
\end{enumerate}

At this point, for the $j^{\text{th}}$ surface and so for the $j^{\text{th}}$ normal direction, the incidence and viewing angles can be computed both for the sunshine and the earthshine contributions, as shown hereafter. The reflectances and surface's areas are indicated as $\rho^{(j)}$ and $A^{(j)}$ respectively.

\begin{table}[htb]
\centering
\begin{tabular}{p{1.4cm} p{6.9cm}} 
Sunshine: & 
    \begin{equation}\label{eq: 3D Sunshine}
    \begin{cases}
    \medskip
    F_{S/C,\odot}^{(j)} = F_{\odot}\dfrac{\rho^{(j)} A^{(j)}}{\pi \norm{\vec{r_o}}^2}\cos\bigl(\theta_i^{(j)}\bigr)\cos\bigl(\theta_r^{(j)}\bigr)\\
    \medskip
    \cos\bigl(\theta_i^{(j)}\bigr)=\dfrac{\vec{\hat{n}_j}\cdot \vec{r_{S/C\rightarrow Sun}}}{\norm{\vec{r_{S/C\rightarrow Sun}}}}\\
    \medskip
    \cos\bigl(\theta_r^{(j)}\bigr)=-\dfrac{\vec{\hat{n}_j}\cdot \vec{r_o}}{\norm{\vec{r_o}}}\\
    j = 1,2,\dots 6\\
    \end{cases}
    \end{equation}\\
 & \\
Earthshine: &
    \begin{equation}\label{eq: 3D Earthshine}
    \begin{cases}
    \medskip
    \psi(\alpha_{\oplus}) = [\sin(\alpha_{\oplus})+(\pi-\alpha_{\oplus})\cos(\alpha_{\oplus})]\\
    \medskip
    F_{\oplus} = \dfrac{2A_{\oplus}R_{\oplus}^2F_{\odot}}{3\pi \norm{\vec{r_o}}^2}\psi(\alpha_{\oplus})\\
    \medskip
    F_{S/C,\oplus}^{(j)} = F_{\oplus}\dfrac{\rho^{(j)} A^{(j)}}{\pi \norm{\vec{r_o}}^2}\cos\bigl(\theta_{i,\oplus}^{(j)}\bigr)\cos\bigl(\theta_r^{(j)}\bigr)\\
    \medskip
    \cos\bigl(\theta_{i,\oplus}^{(j)}\bigr)=-\dfrac{\vec{\hat{n}_j}\cdot \vec{r}}{\norm{\vec{r}}}\\
    \medskip
    \cos\bigl(\theta_r^{(j)}\bigr)=-\dfrac{\vec{\hat{n}_j}\cdot \vec{r_o}}{\norm{\vec{r_o}}}\\
    j = 1,2,\dots 6\\
    \end{cases} 
    \end{equation}
    \\
\end{tabular}
\end{table}

Finally, the visual magnitude of the 3D prism can be estimated as stated in \cref{eq: 3D magnitude}. 

\begin{equation}\label{eq: 3D magnitude}
    m = -26.74 - 2.5\log_{10}\biggl(\dfrac{\sum_{j=1}^6F_{S/C,\odot}^{(j)}+F_{S/C,\oplus}^{(j)}}{F_{\odot}}\biggr)
\end{equation}

Note that this model may be improved by adding all the surfaces that compose the real spacecraft such as solar arrays and antennae. The presence of these items within the model may augment the system fidelity and complexity. Indeed, unusual brightness peaks and flares might be associated with the reflection of specific on-board instruments. The 2D-planar model already discussed in \cref{subsec: 2D plane} can be seen as a proper first-order approximation of their own reflection of light.

\paragraph{Solar Arrays} Solar arrays, in order to exploit as much as possible to incoming sunlight for the generation of needed power, are modelled to be always Sun-pointing. This means that the satellite's solar panels are mounted on the spacecraft by means of 3D joints. The solar array normal vector points towards the direction of the Sun and so its mathematical formulation is given by:

\begin{equation}\label{eq: ns}
    \vec{\hat{n}}_{SA} = \dfrac{\vec{r_{S/C\rightarrow \odot}}}{\norm{\vec{r_{S/C\rightarrow \odot}}}},
\end{equation}

The photometric quantities are computed by following the steps shown in \cref{eq: 3D Sunshine,eq: 3D Earthshine}. Finally, the radiant flux density given by the solar arrays is summed to the spacecraft ones to retrieve the visual magnitude of the prism-solar panel complex. 

\paragraph{Antenna} The antenna contribution is added to the rest of the spacecraft as explained for solar arrays. Looking at the normal vector, the antenna can be modelled to be generically oriented in the space. In fact, small antennae usually change their attitude during the flight in order to optimise the data transmission with the closest ground station. In this context, two scenarios are analysed. They will be deeply discussed in \cref{sec: Antenna} where the operative and \virgolette{off-nominal} conditions are taken into account. On one hand, the first condition is represented by the \emph{observer-pointing} antenna in which:

\begin{equation}\label{eq: Ant nominal}
    \vec{\hat{n}}_{Ant} = -\dfrac{\vec{r_o}}{\norm{\vec{r_o}}}
\end{equation}

On the other hand, in the second scenario, the antenna points in the along-track direction. Therefore, the normal vector is given by the formula shown in \cref{eq: t,eq: n2}.

The corresponding photometric quantities and apparent magnitude can be retrieved as expressed in \cref{eq: 3D Sunshine,eq: 3D Earthshine,eq: 3D magnitude} and in this latter the contribution of the antenna is summed up to the cube and eventual solar arrays (\cref{eq: final m}).

\begin{equation}\label{eq: final m}
\begin{split}
    m = -26.74-{2.5}\log_{10}\biggl[\dfrac{(\sum_{j=1}^6F_{S/C,\odot}^{(j)}+F_{S/C,\oplus}^{(j)})}{F_{\odot}}+\\+\dfrac{2(F_{SA,\odot}+F_{SA,\oplus})+(F_{Ant,\odot}+F_{Ant,\oplus})}{F_{\odot}}\biggr]
\end{split}
\end{equation}


\section{Acquisition and analysis of observational data}
\label{Obs_strategy}
All the observational data used and described in the current paper was acquired by the GAL Hassin observatory through the Galhassin Robotic Telescope 1 (GRT1): an OfficinaStellare Ritchey-Chr\'etien telescope, with an aperture of 400 mm, a focal ratio of f/3.8 and a field of view of 82 $\times$ 82 arcminutes. All the images were taken using the r' (Sloan R) filter, a pixel binning of 2x2 (providing a plate scale of 2.44$\arcsec$/pixel) and an exposure time of 1.0 sec. Such a setting resulted optimal to obtain at least one complete image of the satellite streak for each pointing and at the same time to guarantee a good signal-to-noise for the stars in the fields (SNR $\sim$ 25 for stars with r'-mag $\sim$ 10), necessary for the differential photometry procedure used to measure the satellite magnitudes. 

\subsection{Observational strategy}\label{sec: Strategy}

\begin{figure}[htp]
\centering
\subfloat[]{\includegraphics[width = 0.48\textwidth]{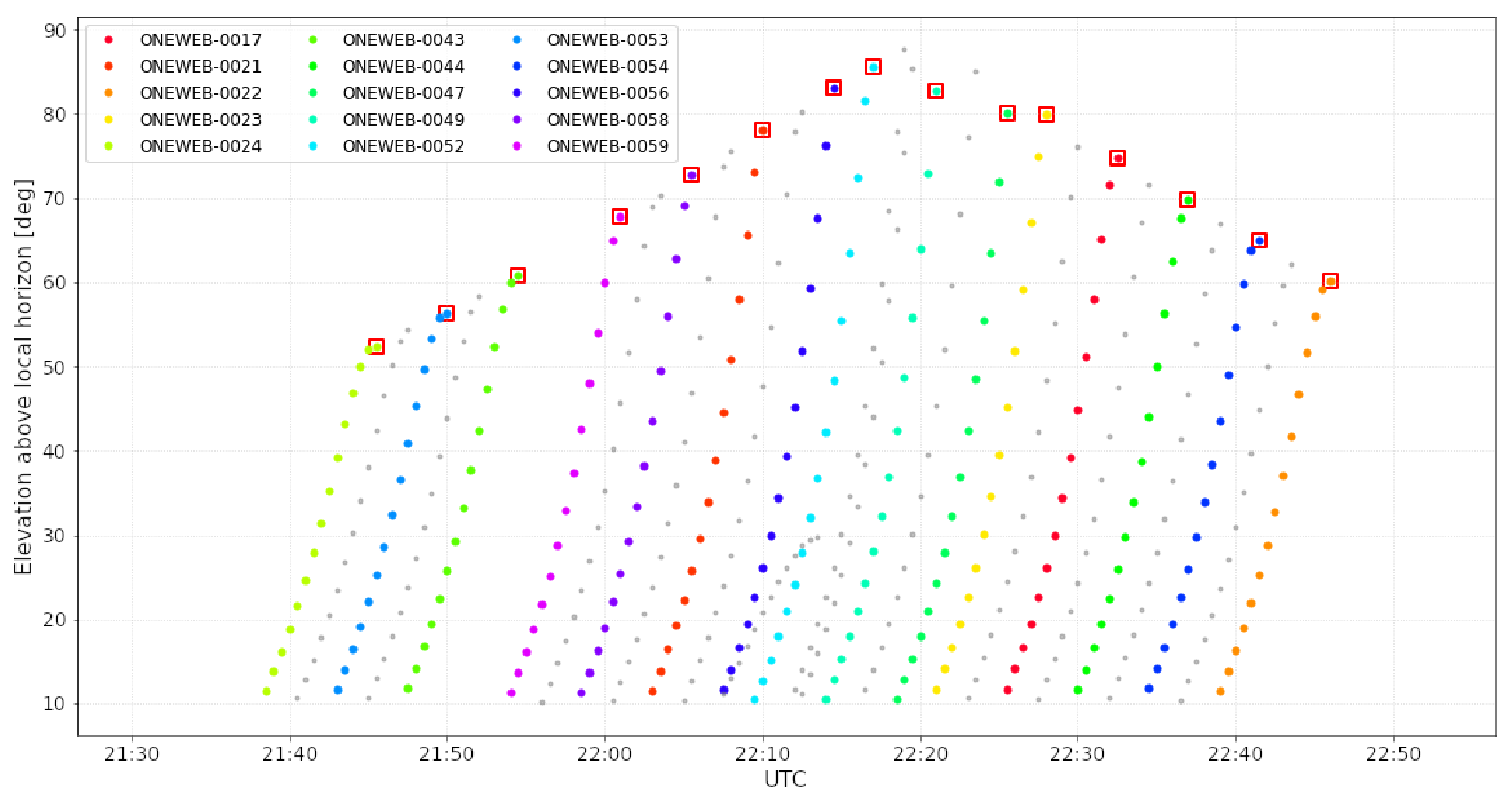}\label{fig: Obs_styartegy Max}}\quad
\subfloat[]{\includegraphics[width = 0.48\textwidth]{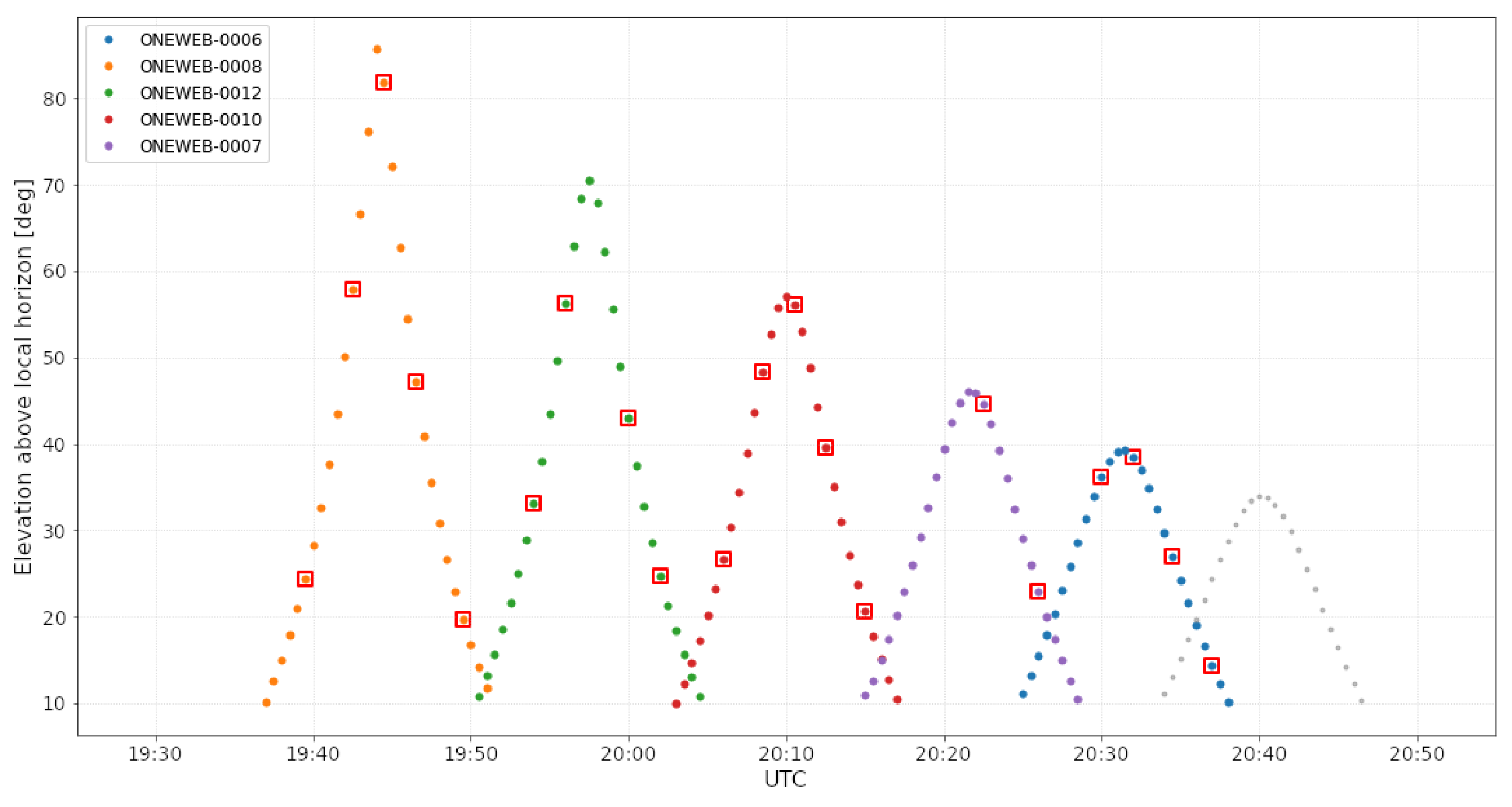}\label{fig: Obs_styartegy Var}}\quad
\caption{Examples of the two strategies adopted for our observational campaign. a) Elevations vs. time for the satellites observed on May 16$^{\rm{th}}$, 2021, in non-eclipsing conditions. The plot depicts the adopted \textit{maximum elevation follow-up} strategy, which consisted in observing and measuring the brightness of the satellites when reaching their maximum elevation, marked by the red boxes. The grey lines mark the position of other satellites in the same train not targeted by our observations. b) Elevations vs. time for the train of satellites observed on July $27^{\rm{th}}$, 2021. In this case the \textit{varying elevation follow-up} was adopted, with each satellite observed multiple times at different elevations in order to map the brightness variation along its path across the sky. Red boxes mark the observed positions for each satellite. All the targeted points were planned also considering the minimum time needed for the telescope slewing, generally amounting to 90 seconds.}
\label{fig: Obs_strategies}
\end{figure}

In order to explore different illumination conditions and hence to map the parameter space in an efficient and complete way, the following two different observational strategies were adopted for the campaign: 

\begin{itemize}
    \item \textit{Maximum elevation follow-up}: each satellite was observed at its maximum elevation above the local horizon. A train of several satellites was targeted in order to cover a wide range of elevations. This strategy was adopted for the set of May $16^{\rm{th}}$, 2021, and it is illustrated in \cref{fig: Obs_styartegy Max}.
    \item \textit{Varying elevation follow-up}: this strategy targeted a lower number of satellites but multiple times along their path across the sky, in order to map their brightness variation at different elevations and illumination conditions. Such a strategy was adopted for the targets observed on July $27^{\rm{th}}$, October $20^{\rm{th}}$ and $21^{\rm{st}}$, and it is illustrated in \cref{fig: Obs_styartegy Var}.
\end{itemize}

The accurate ephemerides of each satellite were calculated via a series of Python programs  making use of module PyEphem (\url{https://rhodesmill.org/pyephem/}) and of the TLEs provided by the Celestrak website, via the link \url{https://celestrak.com/NORAD/elements/supplemental/oneweb.txt}.

Once on position, the telescope started acquiring a continuous series of 1 sec images until the satellite had completely crossed the field. This approach enabled the acquisition of multiple streak images for a same pointing, especially at low elevations because of the lower apparent sky-projected motion (\cref{fig: C1 - OW}).

\subsection{Data analysis}

The images were first reduced using the proper calibration files and standard procedures (dark and bias subtraction, flat fielding correction) and aligned by using the software Tycho Tracker v.8.7 (\url{https://www.tycho-tracker.com/download}). The frames containing the satellite streaks were finally analyzed with the same software using the differential photometry procedure. The \textit{differential photometry} technique makes use of the known magnitudes of the stars in a field to estimate the magnitude of the unknown source(s) in the same image, by comparing their relative brightness. The catalog of magnitudes used for the reference stars is the ATLAS one (\citet{Tonry2018_ATLAS, Kostov2017_ATLAS}). The ATLAS catalog is an all-sky reference catalog comprised of approximately 1 billion stars to a limiting magnitude of 19.  It incorporates data from PanSTARRS DR1, ATLAS PathFinder, ATLAS re-flattened APASS, SkyMapper DR1, APASS DR9, Tycho-2, and the Yale Bright Star Catalog.  The astrometry is sourced from Gaia DR2.  It was developed by \citet{Tonry2018_ATLAS} and serves as a very robust catalog for photometry. For conducting photometry, it is recommended that the native magnitudes of the chosen catalog be used, which in the case of the ATLAS catalog is the r' (Sloan R) magnitude.
Hence, all the satellite magnitudes reported in this paper are provided in r'-mag units.

\begin{figure}[htp]
\centering
\subfloat[]{\includegraphics[width = 0.48\textwidth]{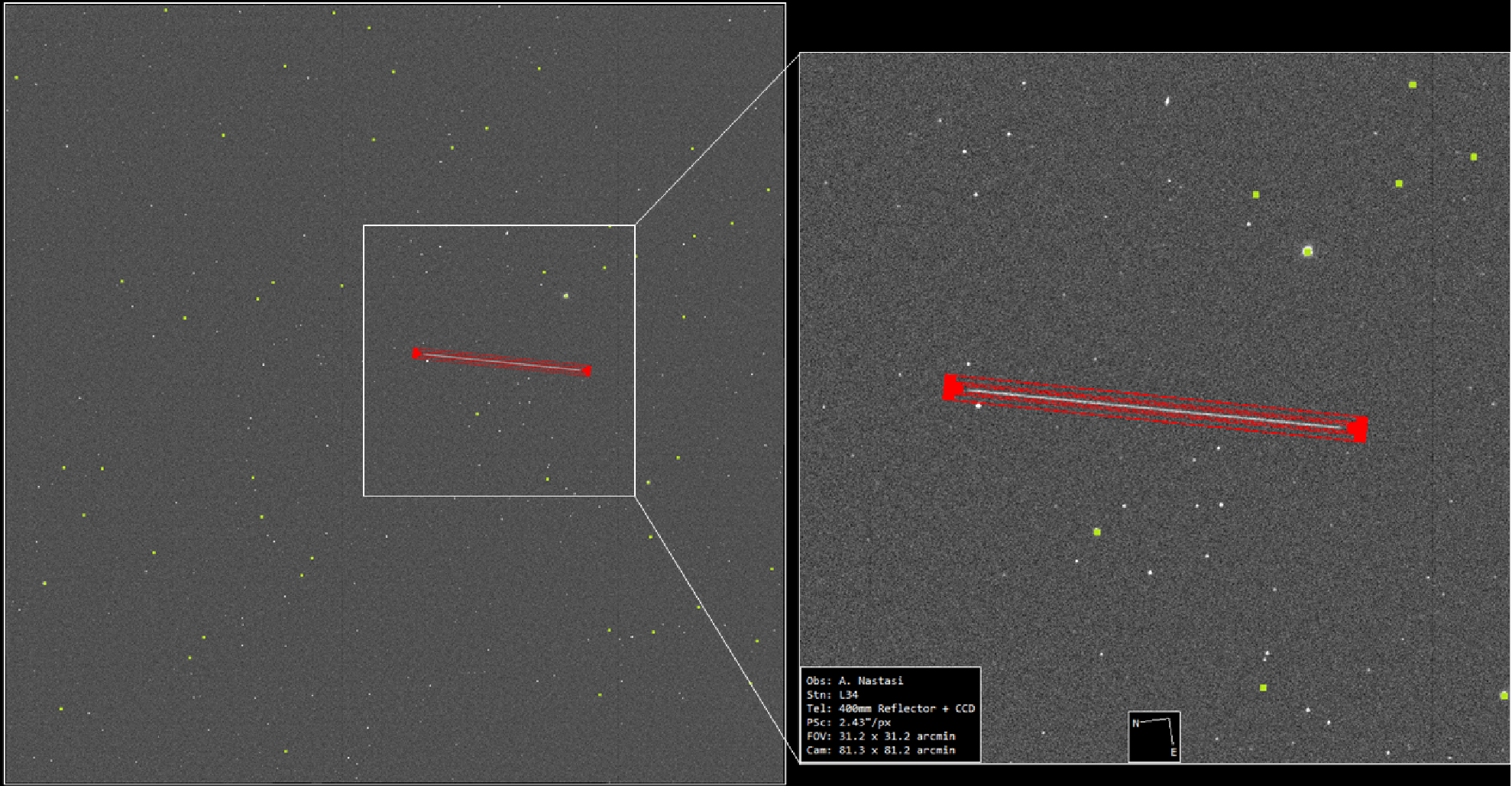}\label{fig: OW131_photometry 1}}\quad
\subfloat[]{\includegraphics[width=0.35\textwidth]{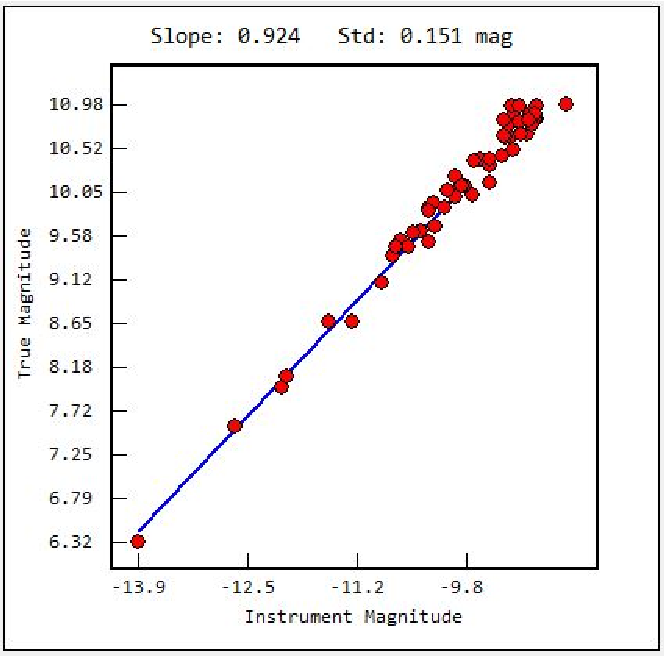}\label{fig: OW131_photometry 2}}\quad
\caption{a) An example frame used for the photometric analysis of a OneWeb satellite, and a closed-up view of the area around its trail. The image recorded the streak left by the satellite ONEWEB-0131 on 20-10-2021 at 19:15:59 UTC, whose magnitude (r'-mag = 7.8) was measured via a rectangular aperture and a set of 51 comparison stars in the field (marked in green). b) Instrumental vs True Magnitude plot for the comparison stars used in the analysis of this specific case. A good linear fit of these quantities guarantees a robust estimate of the magnitudes falling within the same range.}
\label{fig: OW131_photometry}
\end{figure}

The first analysis step consisted in compiling a good set of comparison stars for each imaged field. Stars with mag $\leq$ 11 were selected, as typically measured satellites magnitudes never exceeded the value of 10. Each star was then visually inspected in order to eliminate those ones which resulted less than 50 pixels from the sensor edge, beyond saturation or with one or more visual companion(s) within a radius of 30$\arcsec$. For each image, then, a fixed circular aperture for star photometry was set. The aperture was chosen in order to have a size of about 4 times the full width half maximum (FWHM) of the brightest comparison star. For instance, for a mag 6 star with a FWHM $\sim$ 6.5$\arcsec$, a typical circular aperture of 5 pixels radius (equivalent to 12$\arcsec$), plus a sky annular region of 7 pixels, was set.
Using the selected circular apertures, the software computed the instrumental magnitudes of the comparison stars, which were finally compared with the corresponding (known) catalogue magnitudes to generate a linear fit. This relation was finally used to calibrate the instrumental magnitude of the satellite streak into the corresponding physical quantity (\cref{fig: OW131_photometry 2}).
Properly sized \textit{rectangular apertures} were used to calculate the instrumental and standard magnitudes of satellites. The aperture length was set in order to enclose the entire streak plus a portion of sky of about 30$\arcsec$ per side. The width was typically chosen around 30$\arcsec$, a value well beyond the thickness of all satellite streaks. The rectangular sky region was set to 7 pixels width. These sizes were occasionally slightly modified in order to avoid overlaps with nearby field stars. A visual example of an analyzed frame, with the marked comparisons stars and the rectangular aperture around the satellite streak is reported in \cref{fig: OW131_photometry 1}.

The satellites magnitudes provided by the photometric analysis were finally tested against different choices of rectangular aperture sizes and comparison stars selections, and resulted always consistent within 0.1 mag. Such a value was then set as the typical, albeit conservative, estimate of the uncertainty associated to each measured satellite magnitude.


\section{Model characterization to OneWeb satellites}
\label{sec: Characterization}
The analysis now starts to be specialised on OneWeb satellites. Free parameters (panel areas and reflectivities), involved in the estimation of photometric quantities, shall be identified and fixed. Moreover, the antenna direction can be considered as free variable as well because, usually, antennae feature a steering behaviour which optimises both the download of telemetries and data and the upload of telecommands.


\subsection{Panel areas and reflectivities}\label{sec: Free param}

OneWeb spacecraft is shaped as a main body equipped with two solar arrays mounted on the extremities. Its dimensions can be roughly set to 1 m $\times$ 1 m $\times$ 1.3 m \citep{OW}.
The satellite's body is approximated as a regular prism of dimensions given by the aforementioned reference. Looking at the reflectivity, the value associated with the Multi-Layer Insulator (\emph{MLI}) surrounding the prism is found out in literature. \citet{MLI_rho} described a very detailed analysis carried on the estimation of diffusive reflectivity of \emph{MLI}s and, as average value over the under-analysis spectrum, $\rho_{MLI} = 0.03$ is selected. Focusing on solar arrays, a specific datum is not indicated in literature. Therefore, as first approximation, the dimensions of this latter is decided to be 2 m $\times$ 1.5 m. The reflectivity of solar arrays is discussed by \citet{SA_rho} and, by following the same procedure adopted for the \emph{MLI}, the mean value of this parameter is given by 0.01. Finally, focusing on the steerable antenna, its radius is lower than 150 mm and it is white painted. The reflectivity associated with this bright item is set to 0.65. 
All the parameters that will be exploited inside the analysis are summarised in \cref{tab: Param}. 

\begin{table}[htb]
    \centering
    \caption{Thermo-optical and geometrical properties of the modelled satellite.}
    \label{tab: Param}
    \begin{tabular}{|c|c|c|}
    \hline
    \textbf{Item} & \textbf{Numerical value} & \textbf{Notes}\\
    \hline
    $A_1$ & $1.0\times 1.0$ m$^2$ & \makecell{Nadir-pointing\\ face}\\
    \hline
    $A_2$ & $1.0\times 1.3$ m$^2$ & \makecell{Along-track\\ face}\\
    \hline
    $A_3$ & $1.0\times 1.0$ m$^2$ & \makecell{Opposite face\\with respect to $A_1$}\\
    \hline
    $A_4$ & $1.0\times 1.3$ m$^2$ & \makecell{Opposite face\\ with respect to $A_2$}\\
    \hline
    $A_5$ & $1.0\times 1.3$ m$^2$ & \makecell{Cross-track\\ face}\\
    \hline
    $A_6$ & $1.0\times 1.3$ m$^2$ & \makecell{Opposite face\\ with respect to $A_5$}\\
    \hline
    $A_{SA}$ & $2.0\times 1.5$ m$^2$ & \makecell{Solar\\ arrays}\\
    \hline
    $A_{Ant}$ & $0.15^2\pi$ m$^2$ & \makecell{Gateway\\ antenna (steerable)}\\
    \hline
    $\rho_{MLI}$ & 0.03 & \makecell{Prism\\ faces}\\
    \hline
    $\rho_{SA}$ & 0.01 & \makecell{Solar\\ arrays}\\
    \hline
    $\rho_{Ant}$ & 0.65 & \makecell{Gateway\\ antenna (steerable)}\\
    \hline
    \end{tabular}
\end{table}

\begin{figure}[htp]
    \centering
    \includegraphics[width = 0.4\textwidth]{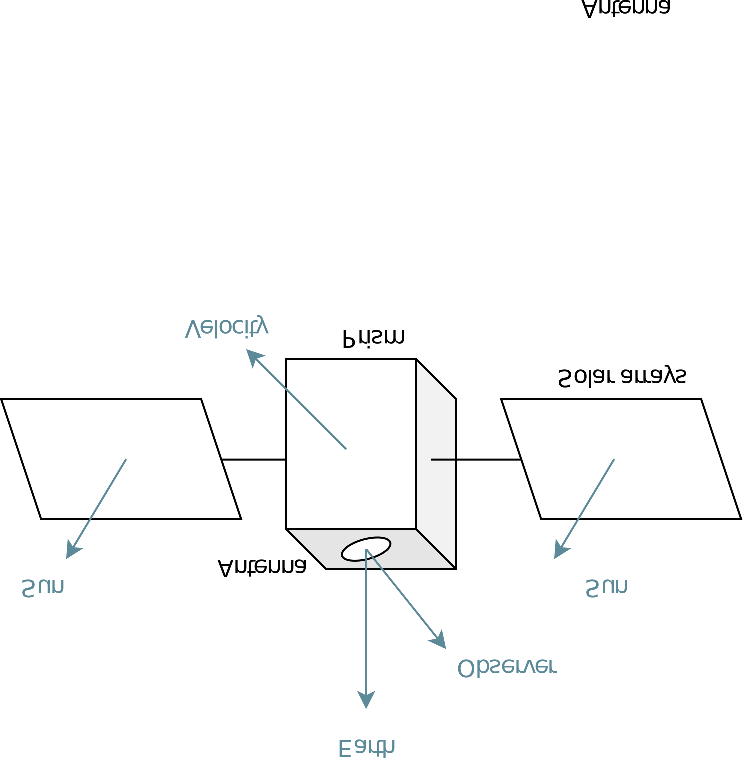}
    \caption{Sketch of the satellite implemented inside the brightness model.}
    \label{fig: Model}
\end{figure}

\begin{figure}[h!]
    \centering
    \includegraphics[width = 0.4\textwidth]{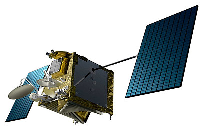}
    \caption{Realistic representation of a OneWeb spacecraft by \citet{OW}.}
    \label{fig: OW SC}
\end{figure}

A very schematic representation of the satellite implemented inside the brightness model is portrayed into \cref{fig: Model}. It is clear that a more realistic sketch of the geometry, in terms of both surface's definition and orientations, could help the model with the computation of the satellite brightness (a realistic picture of OneWeb spacecraft is shown in \cref{fig: OW SC}).

\subsection{Antenna's pointing direction}\label{sec: Antenna}

Antennae of a satellite are fundamental equipment for the transmission of data from and towards the ground stations. In order to minimise losses of link budget and to maximise the performance of the Tracking, TeleMetry and TeleCommand (\emph{TTM\&TC}) subsystem, the misalignment between on-board antenna and ground stations shall be as low as possible. In this analysis, in nominal conditions, the steerable antenna is modelled to be observer-pointing because the closest ground station to GAL Hassin park is located near Palermo. On the other hand, in off-nominal conditions, the antenna does not point in the direction of the ground station. In this case, as a simplification, the antenna is chosen to be aligned with the along-track direction (the direction of velocity for a circular orbit).
In any case, more information about the satellite attitude both in nominal and in off-nominal conditions, could help to optimise the results, becoming closer and closer to observation data provided by the astronomical observatory.


\subsection{Application of the brightness model via observational data}\label{sec: Validation}

Hereafter, three observational campaigns carried out by GAL Hassin observatory are exploited to be compared with model estimations. The first of them was performed on May 16$^{\rm{th}}$, 2021 in which fifteen satellites were observed and nineteen data were retrieved. The strategy adopted for this campaign is represented by the maximum elevation follow-up, already described in \cref{sec: Strategy}. Then, on July 27$^{\rm{th}}$, 2021, five non-operational satellites (nominal orbit but not nominal attitude) were observed and thirty-nine observational sets of information are provided. In this case (and in the following ones), a more complete knowledge of the magnitude evolution is obtained due to the high number of observations per spacecraft (varying elevation follow-up philosophy). The third observational campaign was performed in the night between October 20$^{\rm{th}}$ and October 21$^{\rm{st}}$, 2021 and it is split in two sub-operations: the first of them analysed six satellites (and provided twenty-three photometric data) just after the sunset of the first observational day while the second one looked at nine objects (with thirty-five sets of observational information) before the sunrise of the second day.

\begin{figure*}[htp]
\centering
    \subfloat[][May 16$^{\rm{th}}$, 2021]{\includegraphics[width = 0.45\textwidth]{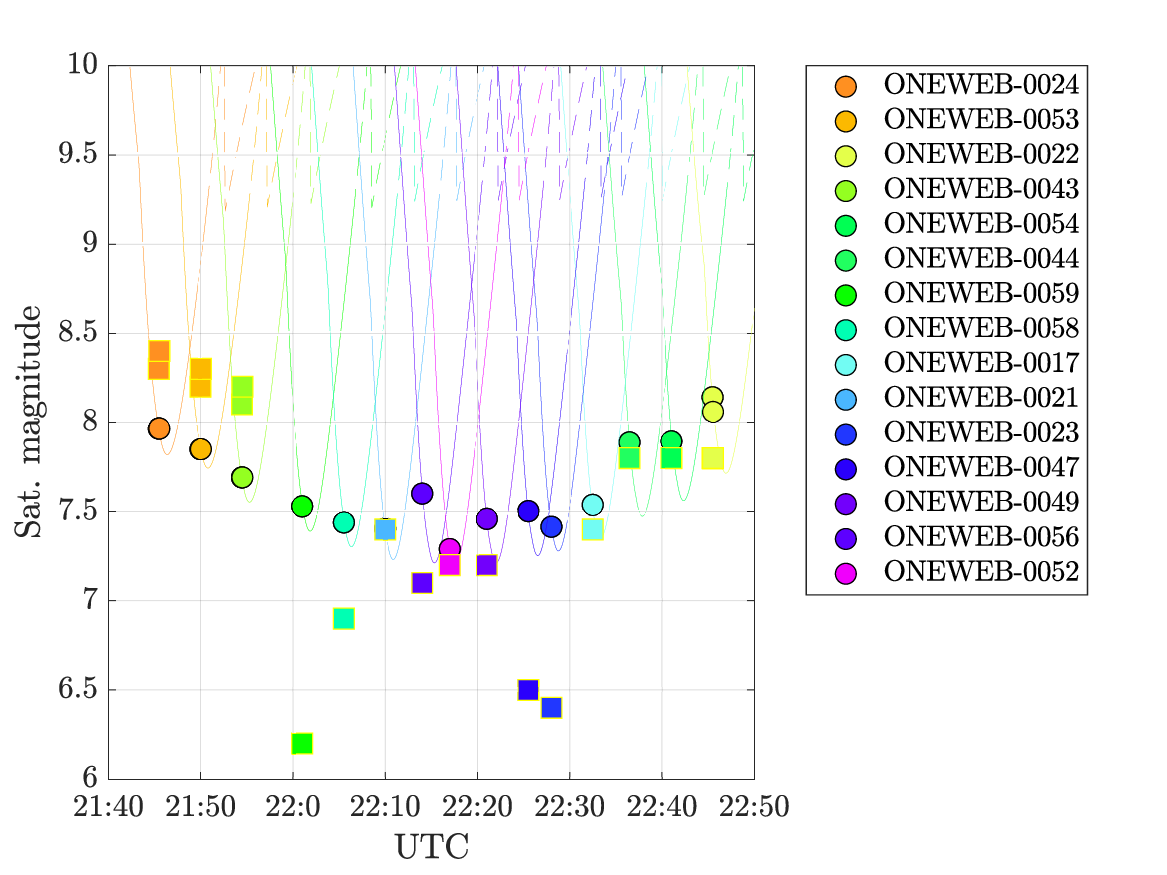}\label{fig: May_Mag_vs_UTC}}\quad
    \subfloat[][July 27$^{\rm{th}}$, 2021]{\includegraphics[width = 0.45\textwidth]{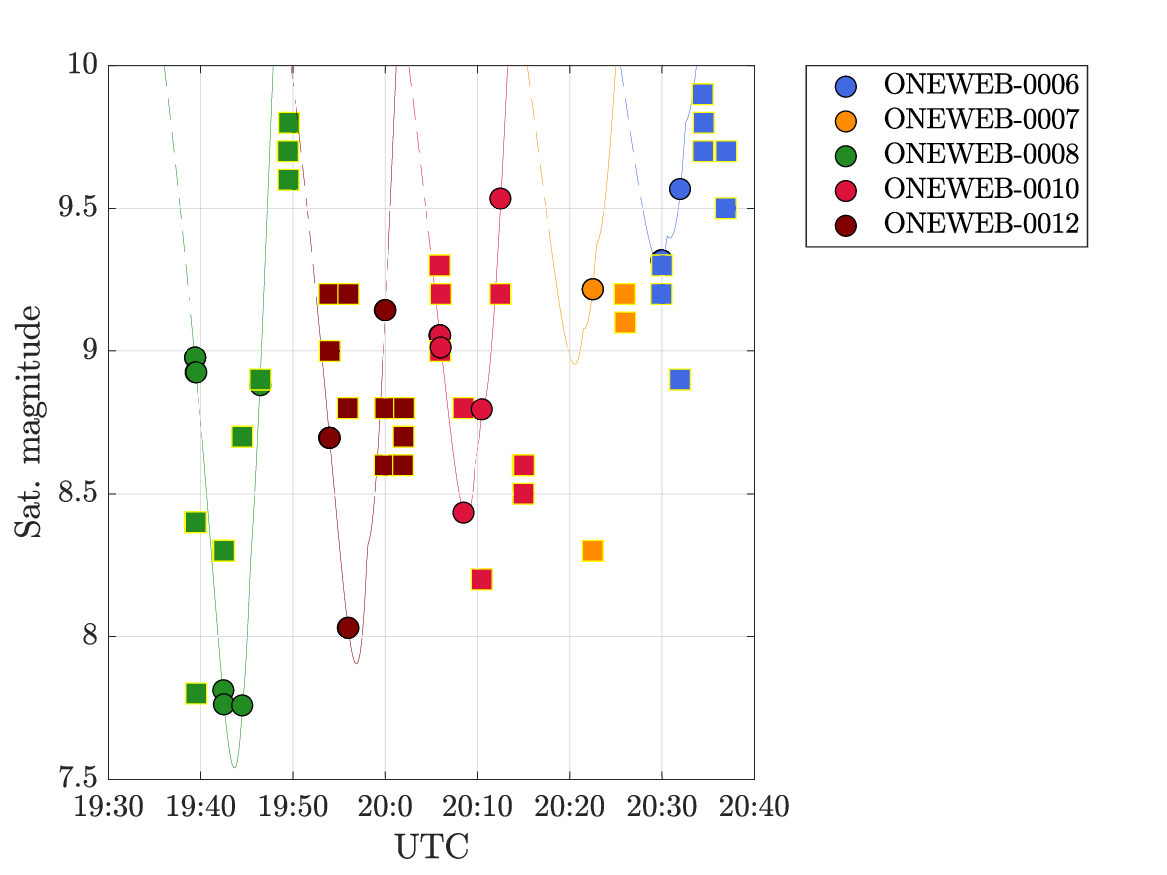}\label{fig: July_Mag_vs_UTC}}\quad
    \subfloat[][October 20$^{\rm{th}}$, 2021]{\includegraphics[width = 0.45\textwidth]{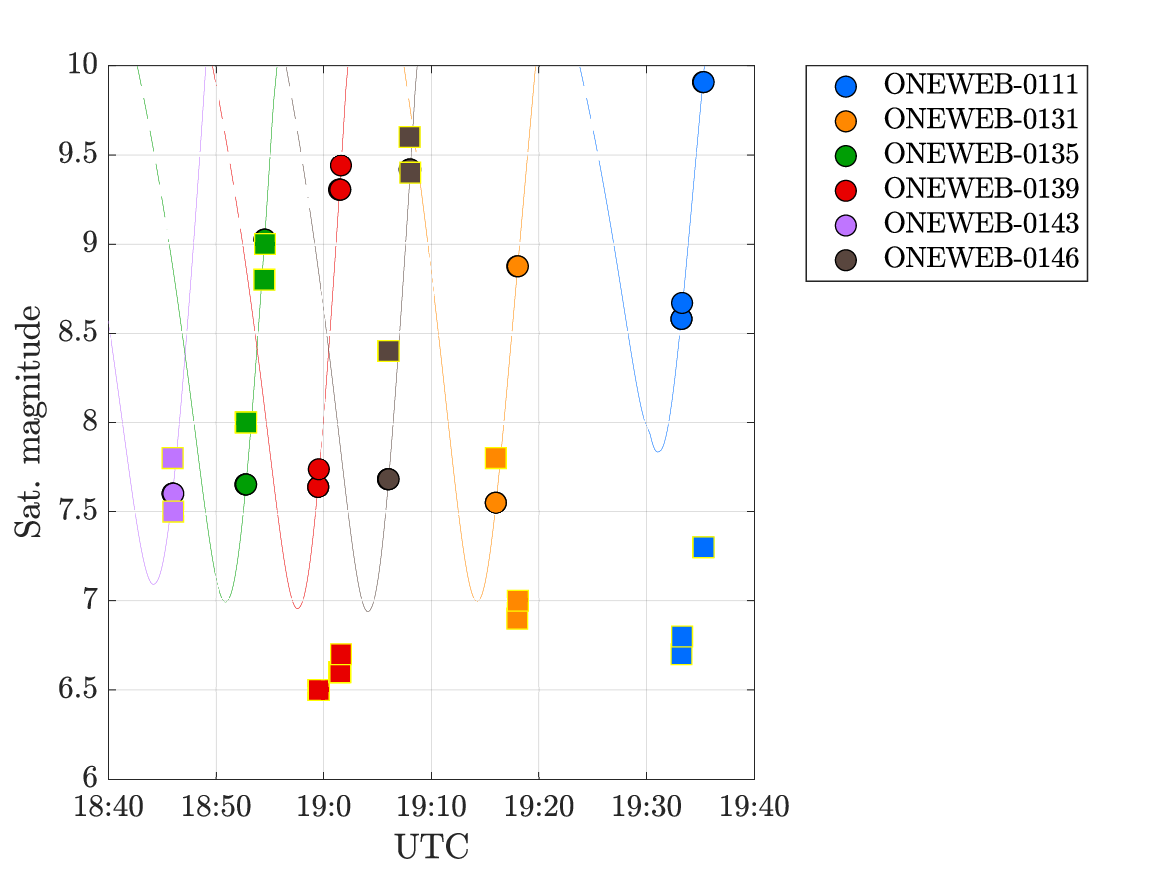}\label{fig: AS_Mag_vs_UTC}}\quad
    \subfloat[][October 21$^{\rm{st}}$, 2021]{\includegraphics[width = 0.45\textwidth]{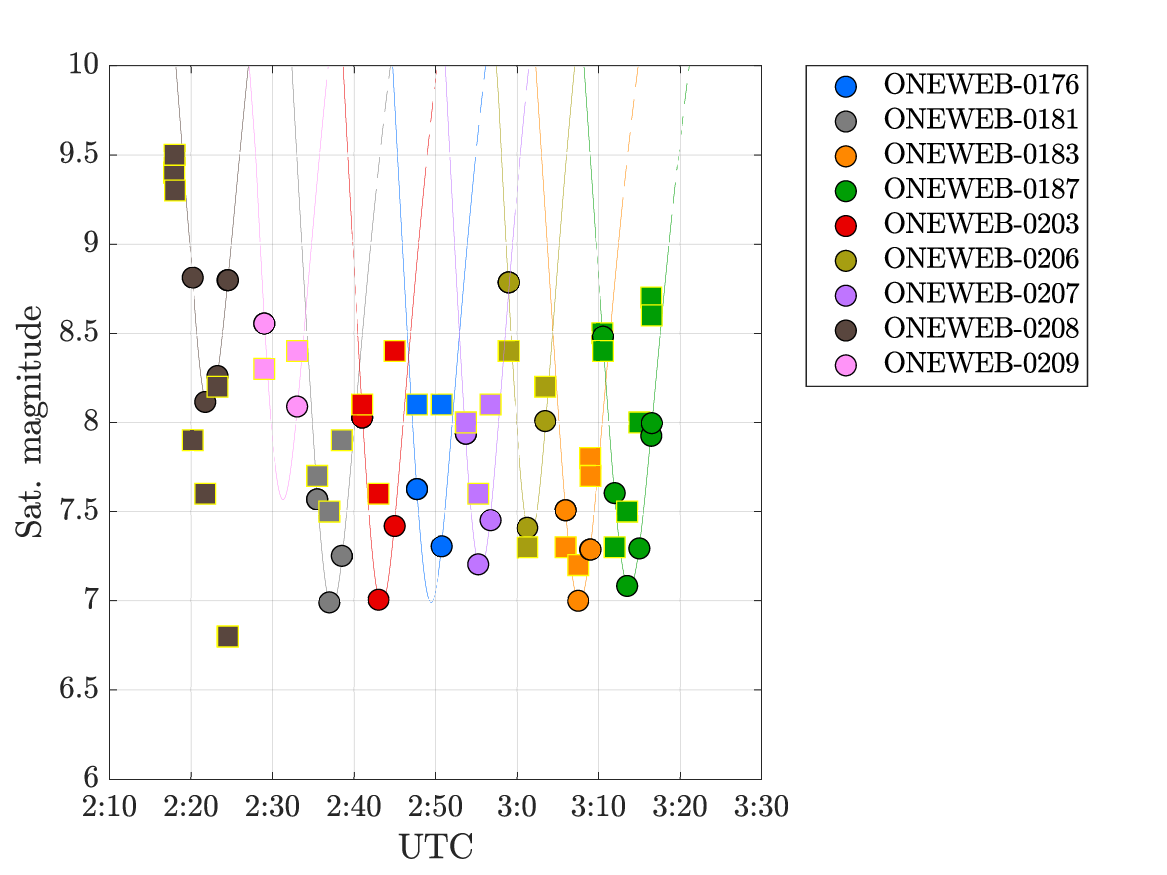}\label{fig: BS_Mag_vs_UTC}}\quad
    \caption{Apparent magnitude versus Coordinated Universal Time (UTC) of the observed OneWeb satellites. Squares represent real data as measured by GAL Hassin observatory, circles refer to the results associated with the implemented brightness model. The lines show the visibility conditions (above 10deg elevation) of the spacecraft (solid line: visible; dashed line: not visible). Values on magnitude axis are reported increasing upward.}
    \label{fig: Mag_vs_UTC}
\end{figure*}

In the following images, the squares represent the real data coming from GAL Hassin analyses (which are shown in \cref{app: Obs_tables}), while the circles depict the model results associated with the implemented brightness model. Moreover, the continuous lines show the satellite in visibility while the dashed tracks portray the shadowing conditions.
\cref{fig: Mag_vs_UTC} shows the magnitude evolution in time for the observational campaigns. All of them show a good approximation of the apparent magnitude behaviour, focusing the attention on a global point of view. On the other hand, there are some local luminous peaks (note: the lower the magnitude the brighter the object) which are not followed by the model approximations. Limitations and next steps to enhance model accuracy are listed in \cref{sec: Conclusions}. These flares are clearly visible in \cref{fig: May_Mag_vs_UTC,fig: AS_Mag_vs_UTC}. Instead, the discrepancies between modelled and real data shown in \cref{fig: July_Mag_vs_UTC} (let's recall this campaign observed non-operational satellites) are strongly associated with the bad knowledge about the item (especially antenna) attitude. Even if it is modelled to be in the along-side direction, its real orientation is actually unknown. Therefore, this campaign will be not considered in the following analysis (\cref{sec: Prediction}), which aim is the prediction of the satellites' brightness.

\begin{figure*}[htp]
\centering
    \subfloat[][May 16$^{\rm{th}}$, 2021]{\includegraphics[width = 0.45\textwidth]{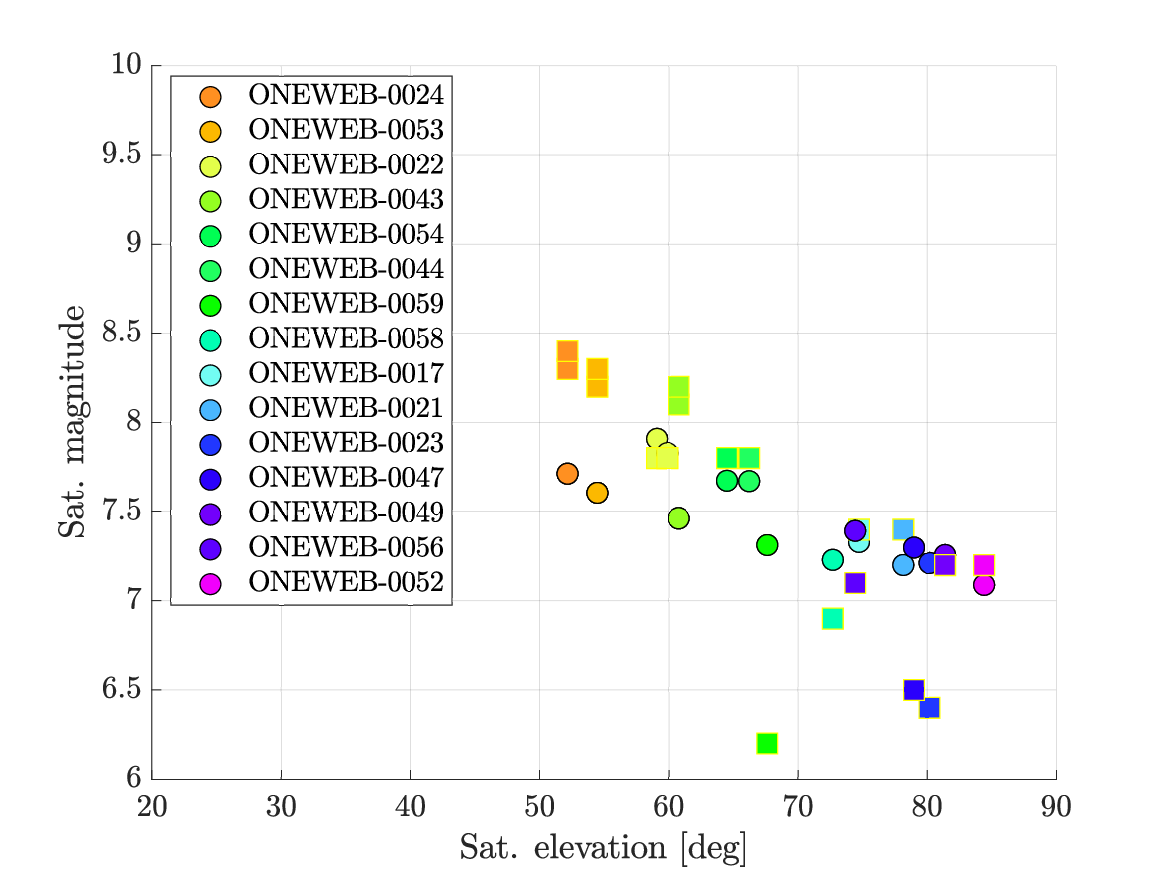}\label{fig: May_Mag_vs_El}}\quad
    \subfloat[][July 27$^{\rm{th}}$, 2021]{\includegraphics[width = 0.45\textwidth]{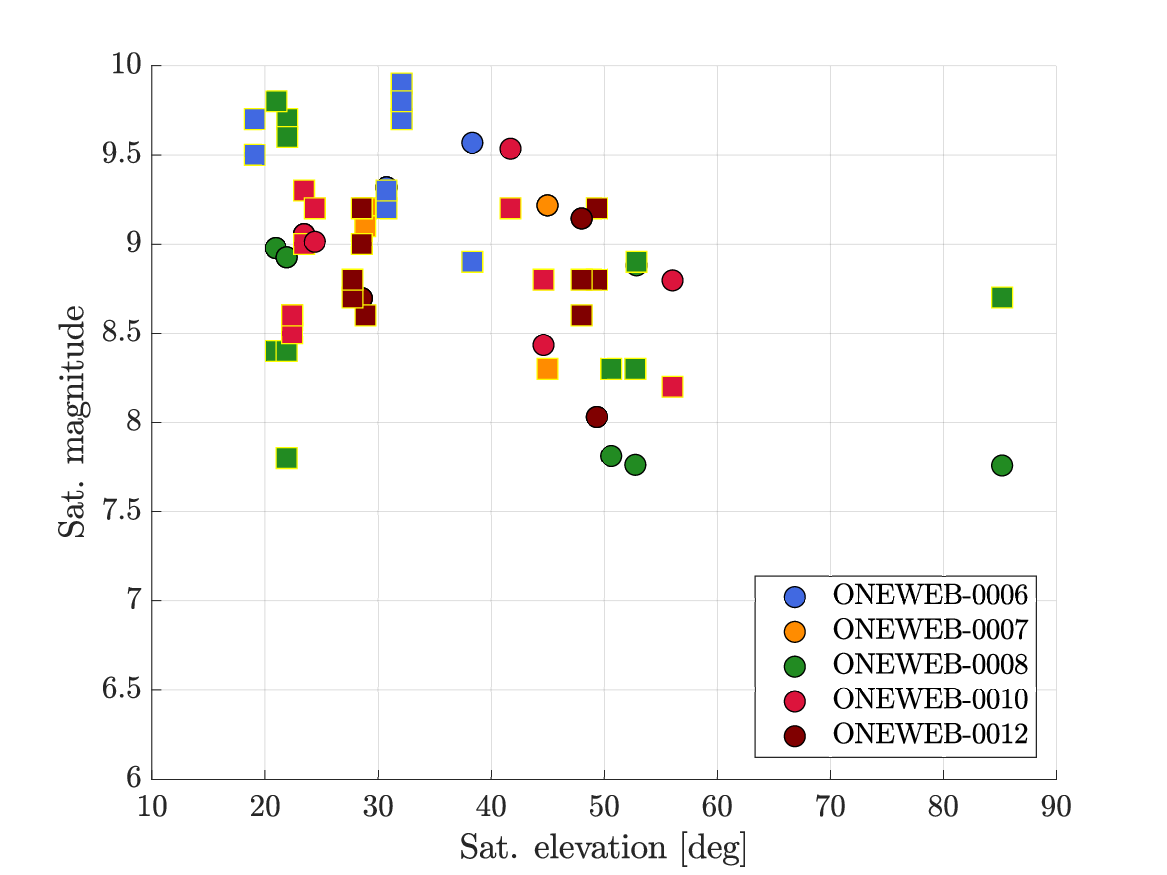}\label{fig: July_Mag_vs_El}}\quad
    \subfloat[][October 20$^{\rm{th}}$, 2021]{\includegraphics[width = 0.45\textwidth]{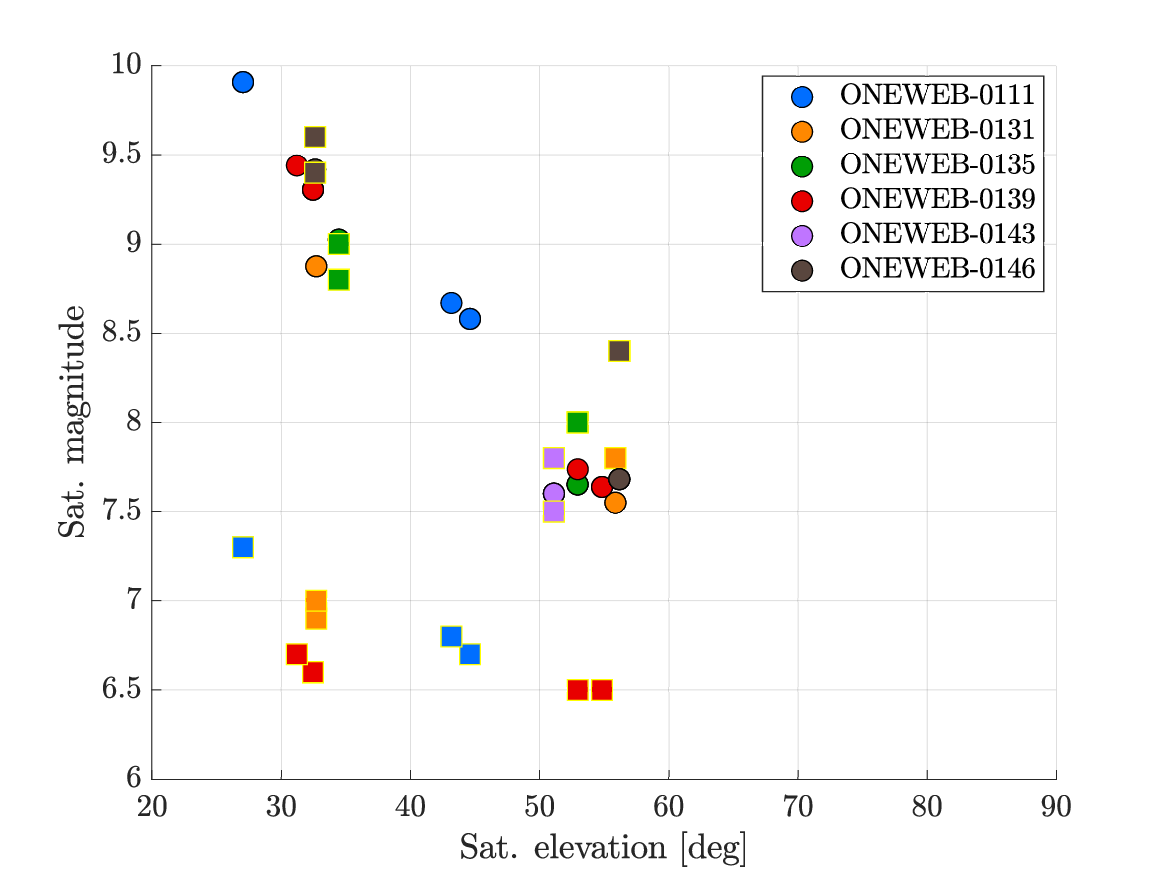}\label{fig: AS_Mag_vs_El}}\quad
    \subfloat[][October 21$^{\rm{st}}$, 2021]{\includegraphics[width = 0.45\textwidth]{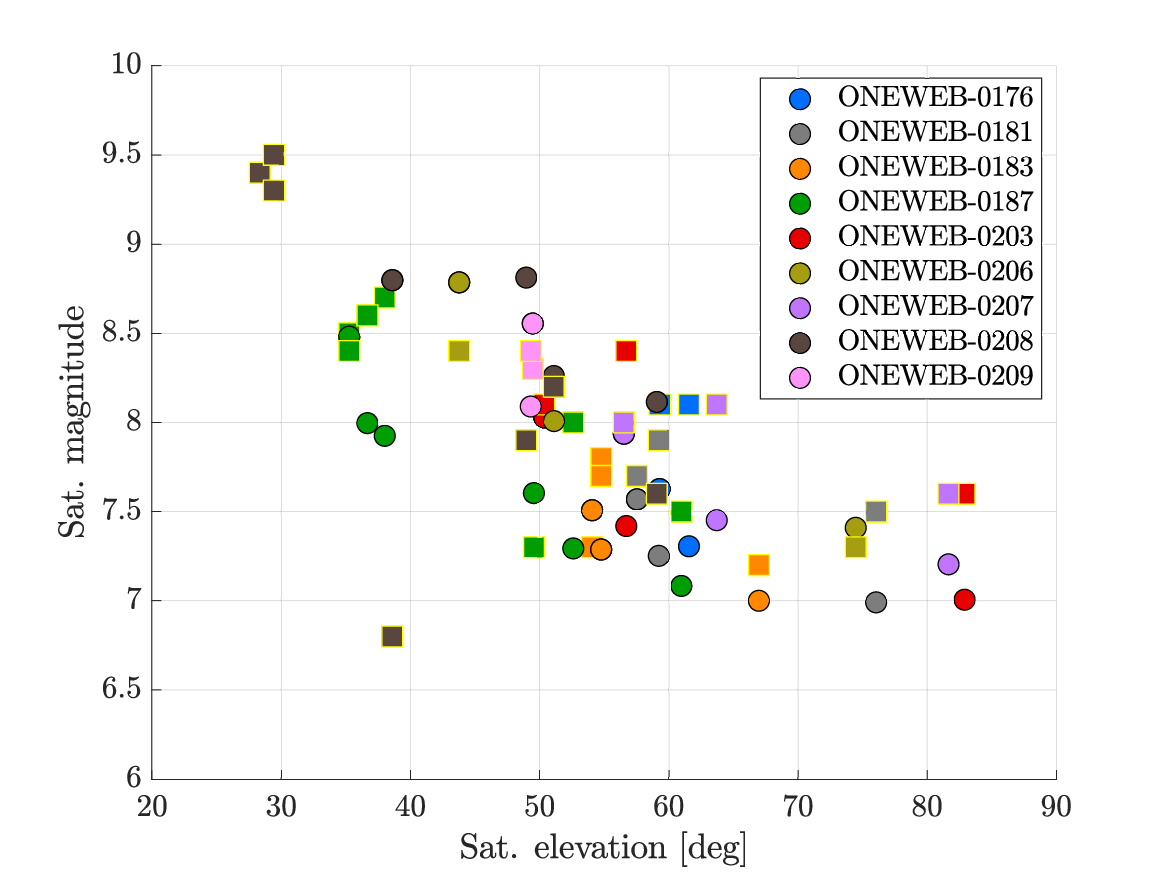}\label{fig: BS_Mag_vs_El}}\quad
    \caption{Apparent magnitude versus Elevation of the observed OneWeb satellites. Squares represent real data as measured by GAL Hassin observatory, circles refer to the results associated with the implemented brightness model. Values on magnitude axis are reported increasing upward.}
    \label{fig: Mag_vs_El}
\end{figure*}

\cref{fig: Mag_vs_El} represents the magnitude variations with respect to the spacecraft elevation. Also in this case it is possible to appreciate a good quality achieved by the brightness model but luminous peaks are not reproduced.

To summarize the results achieved by this comparison test, \cref{tab: Stats} has been formulated. Simple statistical data are provided which can be seen as an initial evaluation of the model effectiveness. Mean values and standard deviations of the differences (in absolute value) between model results and observational data are listed in this table.

\begin{table}[htp]
    \centering
    \begin{threeparttable}
    \caption{Statistical data of differences between model and reality.}
    \label{tab: Stats}
    \begin{tabular}{|c|c|c|}
    \hline
    \makecell{\textbf{Observational}\\\textbf{Campaign}} & \makecell{\textbf{Mean}\\\textbf{value}} & \textbf{STD deviation}\\
    \hline
    May 16$^{\rm{th}}$ & 0.43 & 0.34\\
    July 27$^{\rm{th}}$ & 0.95 & 0.74\\
    \multirow{2}{*}{October 20$^{\rm{th}}$} & 1.27 & 1.04\\
    & 0.28\tnote{$\dagger$} & 0.24\tnote{$\dagger$}\\
    October 21$^{\rm{st}}$ & 0.52 & 0.45\\
    \hline
    \end{tabular}
    \begin{tablenotes}
    \item[$\dagger$]{Without flares.}
    \end{tablenotes}
    \end{threeparttable}
\end{table}

\begin{figure}[htp]
    \centering
    \includegraphics[width = 0.5\textwidth]{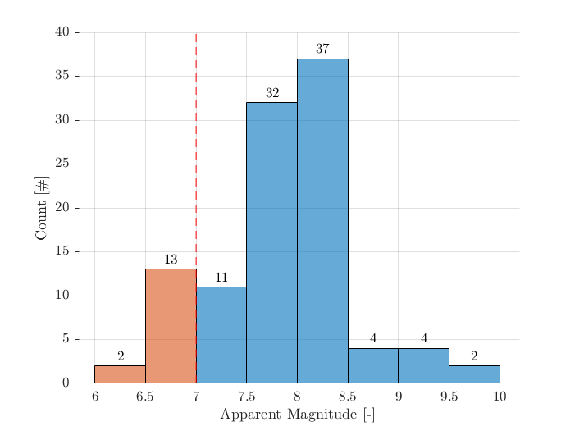}
    \caption{Observation data histogram and count. Dashed red line shows the limit of objects visible to naked eye (7). Bins at magnitude lower than limit are colored in orange.}
    \label{fig: histogram}
\end{figure}

Lastly, the following histogram (\cref{fig: histogram}) shows the observation data count in apparent magnitude ranges between 6 and 10, highlighting a relative small set of OneWeb satellites visible to the naked eye (apparent magnitude $<7$, according to SATCON1 recommendation).

\clearpage

\clearpage


\clearpage
\section{Model validation: brightness predictions}
\label{sec: Prediction}
\begin{figure}[htp]
\centering
\subfloat[][Apparent magnitude versus \emph{UTC}.]{\includegraphics[width = 0.48\textwidth]{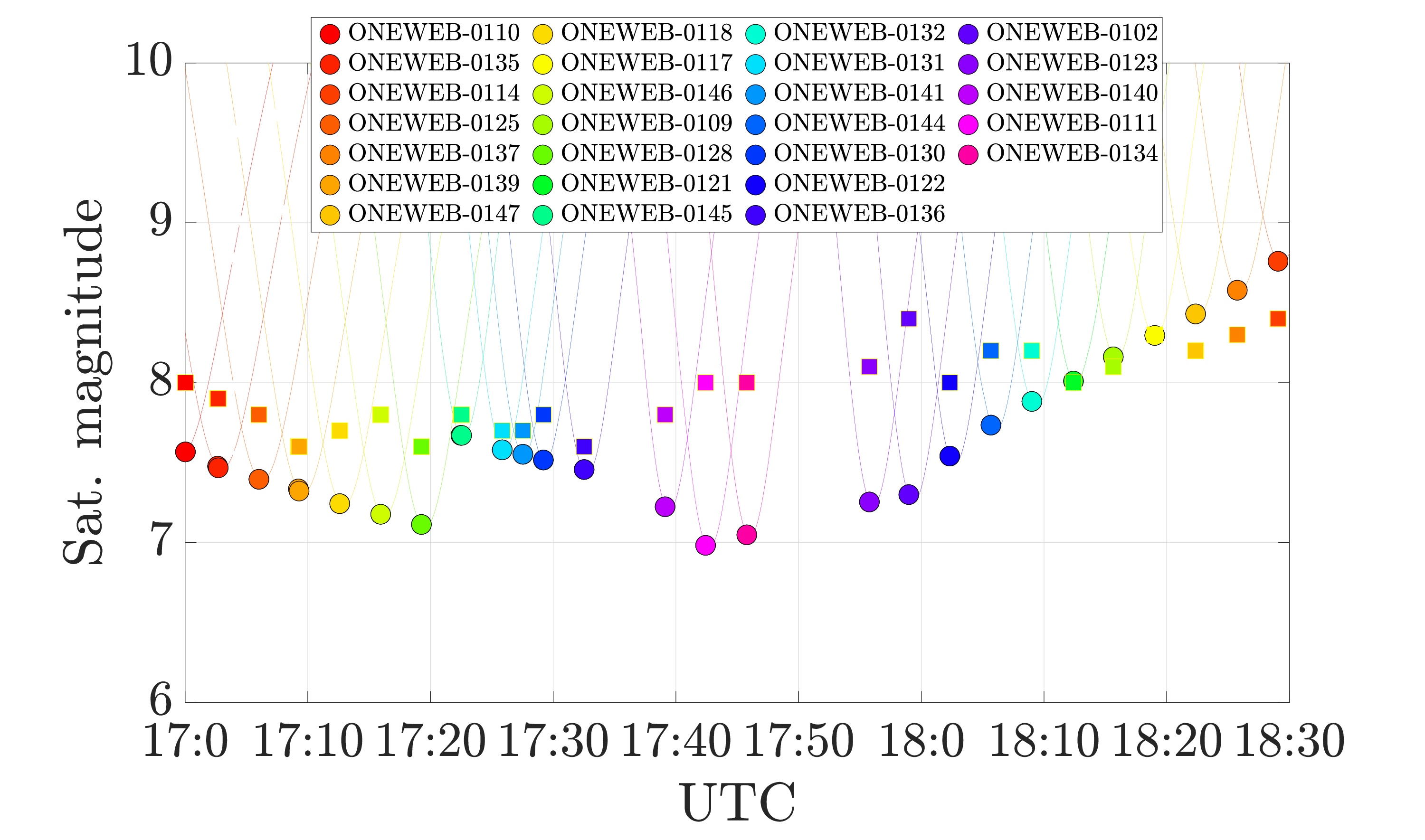}\label{fig: Nov_Mag_vs_UTC}}\quad
\subfloat[][Apparent magnitude versus Elevation.]{\includegraphics[width = 0.48\textwidth]{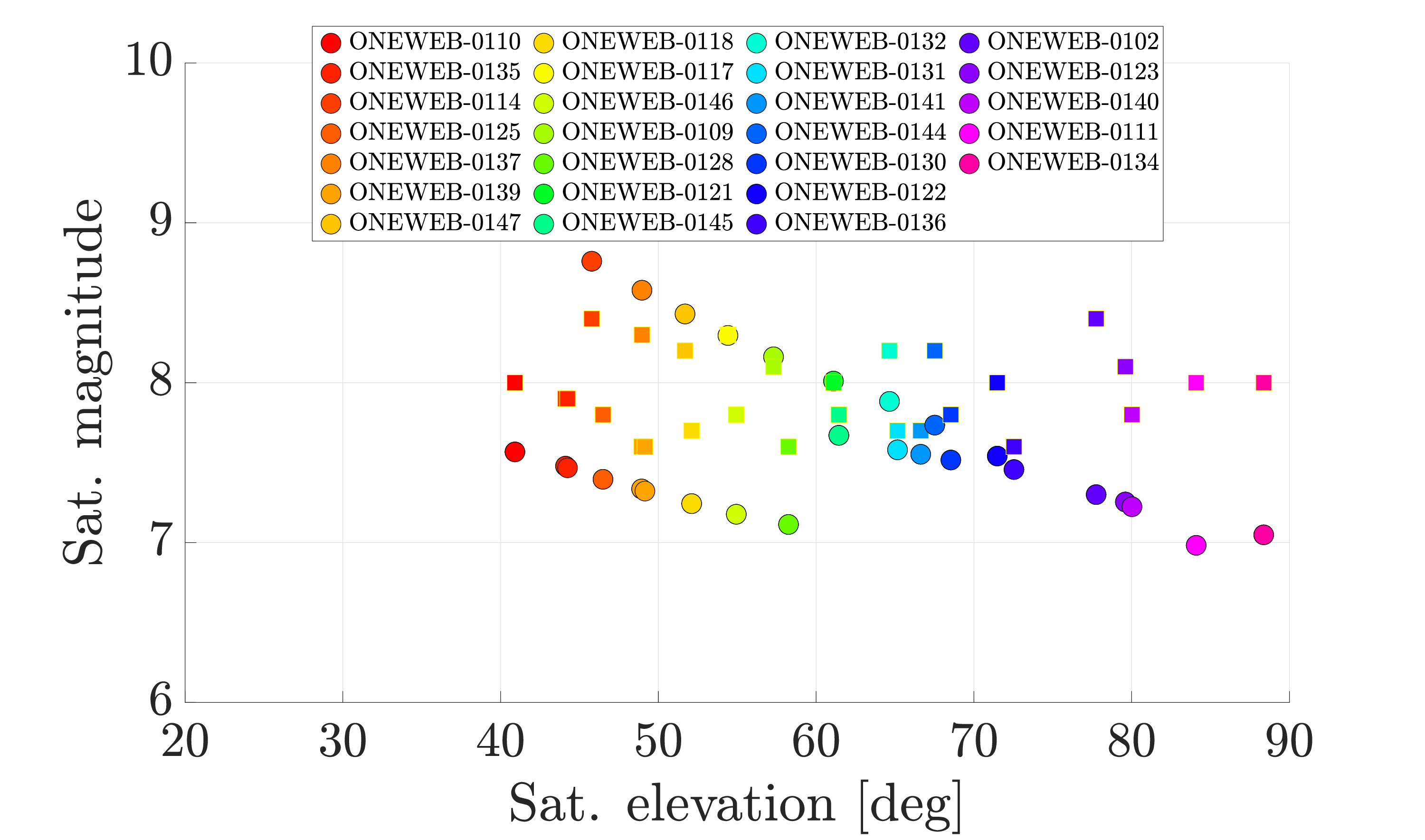}\label{fig: Nov_Mag_vs_El}}\quad
\subfloat[][Differences between predictions and observations.]{\includegraphics[width = 0.48\textwidth]{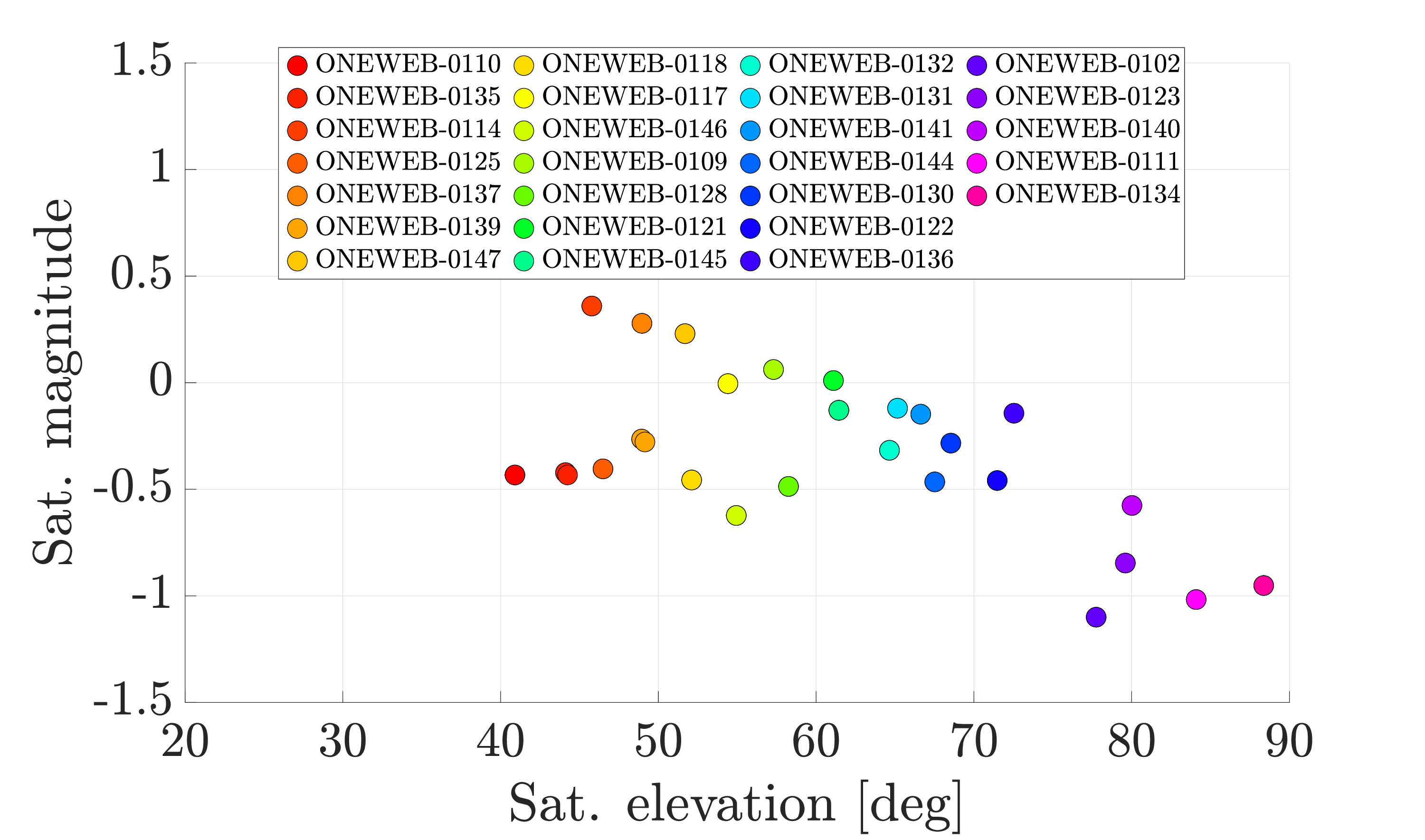}\label{fig: Nov_Diff}}\quad
\caption{November 6$^{\rm{th}}$, 2021 - Predicted magnitude versus Observations of OneWeb satellites. Squares represent real data as measured by GAL Hassin observatory, circles refer to predictions associated with the implemented brightness model, characterised as shown in \cref{sec: Characterization}. The lines - in \cref{fig: Nov_Mag_vs_UTC} - show the visibility conditions (above 10deg elevation) of the spacecraft (solid line: visible; dashed line: not visible). Values on magnitude axis are reported increasing upward.}
\label{fig: Nov}
\end{figure}

This section will show the predictive potential of the model above described. The brightness analysis follows the same steps shown in \cref{sec: Validation} but, in this case, the apparent magnitudes are \emph{predicted} before the measurements executed by the astronomers. This last step could be seen as a final fundamental proof to consolidate the model effectiveness and its own global validity.
The campaign performed by GAL Hassin observatory was carried out on November 6$^{\rm{th}}$, 2021 and twenty-six satellites were spotted, by adopting the maximum elevation follow-up strategy. All the observation data corresponding to this campaign can be found in \cref{app: Obs_tables}.
Let's recall that, in the following pictures (\cref{fig: Nov}), the circles represent the model results while the squares give the correspondent observed magnitude.
\cref{fig: Nov_Mag_vs_UTC} gives a visual representation of the apparent magnitude variation in time, while \cref{fig: Nov_Mag_vs_El} provides the brightness evolution with satellite's elevation. In these images, both limits and strengths associated with the model are clearly visible. Indeed, as already shown in \cref{sec: Validation}, the model is able to globally follow the observations, giving the generic shape of the magnitude evolution, even if local instant variations are not reproduced. The predicted magnitude behaviour appears much more regular than the observed one and this feature leads to a higher discrepancy especially at high elevations, in which the difference between estimations and measurements reaches the unity. Nevertheless, in the other cases, the deviation from reality is bounded between $\pm 1$, as shown in \cref{fig: Nov_Diff}, which is totally consistent with the ranges pointed out during the discussion about model characterization. Finally, mean value and standard deviation of differences between predictions and observations (absolute value) are reported in \cref{tab: Nov}. It is worth to notice the coherence with the numbers represented in \cref{tab: Stats}, being the model able to predict, despite its own limitations, the evolution of satellite's brightness.

\begin{table}[htp]
    \centering
    \caption{Statistical data of differences between model and reality - Predictions.}
    \label{tab: Nov}
    \begin{tabular}{|c|c|c|}
    \hline
    \makecell{\textbf{Observational}\\\textbf{Campaign}} & \makecell{\textbf{Mean}\\\textbf{value}} & \textbf{STD deviation}\\
    \hline
    November 6$^{\rm{th}}$ & 0.39 & 0.28\\
    \hline
    \end{tabular}
\end{table}

\clearpage
\section{Conclusions}
\label{sec: Conclusions}
The development of a brightness model potentially applicable to any shape and dimension is a pure innovation in this field. Indeed, the external surfaces of generic body could be discretised in infinitesimal pieces and, for each of them, \cref{eq: 3D Sunshine,eq: 3D Earthshine} could be applied. Once the normal vectors have been identified for each infinitesimal surface, the incidence and viewing angles can be computed. Knowing their extension and the material characteristics (in terms of reflectivity) the radiant flux density is found out. The apparent magnitude of the body is then computed starting from the flux density retrieved above (\cref{eq: 3D magnitude}).

\cref{sec: Validation,sec: Prediction} have shown the validation of the model developed by Politecnico di Milano. It has been demonstrated the good similarity between computed estimations and real data, particularly if the analysis is focused on the global trend of the brightness evolution. The three observational campaigns carried out at GAL Hassin observatory have highlighted that the model can be taken as a first order method to numerically compute the magnitude that the satellites shine.

Despite the good global approximation achieved by the implemented brightness model, there are some conditions in which flares or luminous peaks are not replicated by the computational estimations. The causes of this failure could be recognised in several factors:
\begin{itemize}
    \item Absence of self-shadowing and self-illumination: these two aspects have not been taken into account in the presented model. However, both the self-illumination (due to the reflection of an item over another subsystem) and the self-shadowing (for instance, a panel that shadows a main body's face) could affect the results provided by the model. The former one (self-illumination) can be a possible cause of flares and peaks because it is characterised by a very instantaneous and local behaviour, strongly dependent on the item orientations. On the other hand, the self-shadowing could help the model to follow real data whenever the results are much brighter than the observations (high-elevation points of the July campaign);
    \item Real attitude of the items: the orientation of the modelled subsystems (main body, solar arrays and antenna) is just an approximation of the real one. The non-operative condition applied for the July campaign has emphasised more this aspect, with a general, large discrepancy between predicted and observed luminosities due to the non-nominal items attitudes. The model could be surely improved by implementing attitudes and orientation closer to the reality;
    \item Not adapted thermo-optical properties: what has been implemented inside the model is a diffusive reflectivity found in literature. Even if a mean value of quasi-standard components is selected, the real one could differ a bit from the average quantity. Deeper awareness in this field minimises the local distance between computations and observations;
    \item Missing information about surface orientations: the satellite has been approximated as a perfect prism but the real satellite's shape differs from the ideal 3D object. The inclinations of the normal vectors to the surfaces directly affects the photometric quantities, by defining different angles with respect to the source and the observer;  
    \item Not perfect knowledge of surface extensions. Panel areas, in fact, are a primary factor inside the expression of the radiant flux density.
\end{itemize}

Possible future works aiming the enhancement of the brightness model implemented and validated inside this study can focus the attention on the topics shown above. Particular attention is required by the last of them being the self-illumination and shadowing fundamental phenomena in the detection of peaks and flares. Another theme that it is critical to investigate is the specular reflection. Indeed, Lambert, Lommel-Seeliger and Area laws analyse only the diffusive behaviour of the reflection. The specular contribution could be another important factor in the generation of local failures of the already implemented model. The cooperative effort provided by specular and diffusive reflections is the best possible approximation of the real phenomenon producing the satellite brightness.

In conclusion, this important result can be thought as a preliminary fundamental step towards the preservation and safeguarding of the astronomer activities and, in general, of the increasingly light-polluted sky.

\section{Acknowledgments}
The research at Politecnico di Milano for the development of the brightness model was founded by the European Union's Horizon Europe research and innovation programme (grant agreement No 101089265 – GREEN SPECIES).

The authors would like to acknowledge Diego Alberto Blanco, Maurizio Vanotti and Dominique Poncet from OneWeb Ltd.


\bibliographystyle{jasr-model5-names}
\biboptions{authoryear}


\begin{appendix}

\section{GAL Hassin observational data}\label{app: Obs_tables}
In this appendix, real data collected by the observatory are shown. The results of each campaign are listed in the following tables in which the spacecraft ID, the observation time (Coordinated Universal Time or \emph{UTC}), satellite elevation and the measured apparent magnitude in r' (Sloan R) band are provided. Data points, as well as TLEs, can be downloaded by using the following link: \href{https://polimi365-my.sharepoint.com/:f:/g/personal/10044670_polimi_it/EhnmXfjjXmdKsw1a3PJn-QoBKtONed2MazxXY1Va1pE9vw?e=yl3ml9}{Brightness model - Database}.

\begin{table}[htb]
    \centering
    \caption{Observational campaign: May 16$^{\rm{th}}$, 2021.}
    \label{tab: C4 - May}
    \begin{tabular}{|c|c|c|c|}
    \hline
    \textbf{Satellite} & \textbf{UTC} & \makecell{\textbf{Elevation}\\\textbf{[deg]}} & \makecell{\textbf{Apparent}\\\textbf{Magnitude}}\\
    \hline
    \textbf{ONEWEB-0017} & 22:32:30.0 & 74.73 & 7.4\\
    \hline
    \textbf{ONEWEB-0021} & 22:10:00.6 & 78.16 & 7.4\\
    \hline
    \multirow{2}{*}{\textbf{ONEWEB-0022}} & 22:45:29.3 & 59.10 & 7.8\\
    & 22:45:32.2 & 59.30 & 7.8\\
    \hline
    \textbf{ONEWEB-0023} & 22:28:01.6 & 79.95 & 6.4\\
    \hline
    \multirow{2}{*}{\textbf{ONEWEB-0024}} & 21:45:28.3 & 52.36 & 8.3\\
    & 21:45:31.7 & 52.31 & 8.4\\
    \hline
    \multirow{2}{*}{\textbf{ONEWEB-0043}} & 21:54:28.9 & 60.79 & 8.1\\
    & 21:54:31.6 & 60.75 & 8.2\\
    \hline
    \textbf{ONEWEB-0044} & 22:36:29.2 & 67.50 & 7.8\\
    \hline
    \textbf{ONEWEB-0047} & 22:25:31.2 & 80.34 & 6.5\\
    \hline
    \textbf{ONEWEB-0049} & 22:21:01.3 & 83.05 & 7.2\\
    \hline
    \textbf{ONEWEB-0052} & 22:17:00.5 & 85.39 & 7.2\\
    \hline
    \multirow{2}{*}{\textbf{ONEWEB-0053}} & 21:49:59.1 & 56.37 & 8.2\\
    & 21:50:01.9 & 56.32 & 8.3\\
    \hline
    \textbf{ONEWEB-0054} & 22:41:01.4 & 63.84 & 7.8\\
    \hline
    \textbf{ONEWEB-0056} & 22:14:01.4 & 83.20 & 7.1\\
    \hline
    \textbf{ONEWEB-0058} & 22:05:31.7 & 72.74 & 6.9\\
    \hline
    \textbf{ONEWEB-0059} & 22:01:00.3 & 67.68 & 6.2\\
    \hline
    \end{tabular}
\end{table}

\begin{table}[htp]
    \centering
    \caption{Observational campaign: July 27$^{\rm{th}}$, 2021.}
    \label{tab: C4 - July}
    \begin{tabular}{|c|c|c|c|}
    \hline
    \textbf{Satellite} & \textbf{UTC} & \makecell{\textbf{Elevation}\\\textbf{[deg]}} & \makecell{\textbf{Apparent}\\\textbf{Magnitude}}\\
    \hline
    \multirow{8}{*}{\textbf{ONEWEB-0006}} & 20:29:56.7 & 31.36 & 9.2\\
    & 20:30:00.1 & 31.65 & 9.3\\
    & 20:31:57.7 & 38.54 & 8.9\\
    & 20:34:25.0 & 31.71 & 9.9\\
    & 20:34:28.7 & 31.38 & 9.7\\
    & 20:34:32.3 & 31.07 & 9.8\\
    & 20:36:55.0 & 18.51 & 9.5\\
    & 20:36:58.0 & 18.22 & 9.7\\
    \hline
    \multirow{3}{*}{\textbf{ONEWEB-0007}} & 20:22:30.7 & 45.41 & 8.3\\
    & 20:25:57.7 & 28.06 & 9.2\\
    & 20:26:01.3 & 27.68 & 9.1\\
    \hline
    \multirow{10}{*}{\textbf{ONEWEB-0008}} & 19:39:24.7 & 21.51 & 8.4\\
    & 19:39:28.1 & 21.89 & 8.4\\
    & 19:39:31.7 & 22.27 & 7.8\\
    & 19:42:27.9 & 52.27 & 8.3\\
    & 19:42:31.7 & 53.24 & 8.3\\
    & 19:44:30.6 & 85.15 & 8.7\\
    & 19:46:28.8 & 51.77 & 8.9\\
    & 19:49:27.9 & 21.87 & 9.7\\
    & 19:49:31.3 & 21.50 & 9.6\\
    & 19:49:34.6 & 21.13 & 9.8\\
    \hline
    \multirow{9}{*}{\textbf{ONEWEB-0010}} & 20:05:54.4 & 23.31 & 9.2\\
    & 20:05:57.8 & 23.68 & 9.0\\
    & 20:06:01.2 & 24.05 & 8.8\\
    & 20:08:29.8 & 44.31 & 9.2\\
    & 20:10:28.4 & 55.91 & 8.6\\
    & 20:12:29.3 & 42.39 & 8.8\\
    & 20:14:55.9 & 23.02 & 8.6\\
    & 20:14:59.3 & 22.67 & 8.7\\
    & 20:15:02.6 & 22.31 & 8.8\\
    \hline
    \multirow{9}{*}{\textbf{ONEWEB-0012}} & 19:53:56.8 & 28.30 & 9.2\\
    & 19:54:00.2 & 28.75 & 9.0\\
    & 19:55:57.6 & 48.75 & 8.8\\
    & 19:56:01.0 & 48.45 & 9.2\\
    & 19:59:58.4 & 48.63 & 8.6\\
    & 20:00:01.7 & 47.95 & 8.8\\
    & 20:01:56.6 & 28.67 & 8.6\\
    & 20:01:59.9 & 28.22 & 8.7\\
    & 20:02:03.3 & 27.79 & 8.8\\
    \hline
    \end{tabular}
\end{table}

\begin{table}[htp]
    \centering
    \caption{Observational campaign: October 20$^{\rm{th}}$, 2021.}
    \label{tab: C4 - AS}
    \begin{tabular}{|c|c|c|c|}
    \hline
    \textbf{Satellite} & \textbf{UTC} & \makecell{\textbf{Elevation}\\\textbf{[deg]}} & \makecell{\textbf{Apparent}\\\textbf{Magnitude}}\\
    \hline
    \multirow{4}{*}{\textbf{ONEWEB-0111}} & 19:33:14.8 & 44.36 & 6.7\\
    & 19:33:18.1 & 43.79 & 6.8\\
    & 19:35:14.8 & 26.76 & 7.3\\
    & 19:35:18.3 & 26.35 & 7.3\\
    \hline
    \multirow{3}{*}{\textbf{ONEWEB-0131}} & 19:15:59.9 & 56.08 & 7.8\\
    & 19:17:59.3 & 32.56 & 6.9\\
    & 19:18:02.8 & 32.05 & 7.0\\
    \hline
    \multirow{4}{*}{\textbf{ONEWEB-0135}} & 18:52:44.6 & 53.53 & 8.0\\
    & 18:52:48.1 & 52.78 & 8.0\\
    & 18:54:29.6 & 34.08 & 8.8\\
    & 18:54:33.0 & 33.57 & 9.0\\
    \hline
    \multirow{6}{*}{\textbf{ONEWEB-0139}} & 18:59:29.9 & 54.65 & 6.5\\
    & 18:59:33.3 & 53.83 & 6.5\\
    & 19:01:26.1 & 32.35 & 6.6\\
    & 19:01:29.6 & 31.84 & 6.6\\
    & 19:01:33.1 & 31.32 & 6.6\\
    & 19:01:36.7 & 30.80 & 6.7\\
    \hline
    \multirow{2}{*}{\textbf{ONEWEB-0143}} & 18:45:58.4 & 51.44 & 7.8\\
    & 18:46:02.1 & 50.78 & 7.5\\
    \hline
    \multirow{4}{*}{\textbf{ONEWEB-0146}} & 19:05:59.0 & 56.70 & 8.4\\
    & 19:06:02.5 & 55.80 & 8.4\\
    & 19:07:59.3 & 32.58 & 9.6\\
    & 19:08:02.9 & 32.05 & 9.4\\
    \hline
    \end{tabular}
\end{table}

\begin{table}[htp]
    \centering
    \caption{Observational campaign: October 21$^{\rm{st}}$, 2021.}
    \label{tab: C4- BS}
    \begin{tabular}{|c|c|c|c|}
    \hline
    \textbf{Satellite} & \textbf{UTC} & \makecell{\textbf{Elevation}\\\textbf{[deg]}} & \makecell{\textbf{Apparent}\\\textbf{Magnitude}}\\
    \hline
    \multirow{2}{*}{\textbf{ONEWEB-0176}} & 02:47:44.5 & 62.21 & 8.1\\
    & 02:50:46.0 & 59.63 & 8.1\\
    \hline
    \multirow{3}{*}{\textbf{ONEWEB-0181}} & 02:35:29.6 & 58.50 & 7.7\\
    & 02:36:59.1 & 76.04 & 7.5\\
    & 02:38:31.0 & 58.54 & 7.9\\
    \hline
    \multirow{4}{*}{\textbf{ONEWEB-0183}} & 03:05:58.2 & 56.14 & 7.3\\
    & 03:07:31.1 & 67.48 & 7.2\\
    & 03:08:58.4 & 53.82 & 7.8\\
    & 03:09:01.7 & 53.14 & 7.7\\
    \hline
    \multirow{7}{*}{\textbf{ONEWEB-0187}} & 03:10:29.1 & 36.37 & 8.5\\
    & 03:10:33.3 & 37.03 & 8.4\\
    & 03:11:59.5 & 51.47 & 7.3\\
    & 03:13:31.3 & 61.22 & 7.5\\
    & 03:15:01.1 & 51.14 & 8.0\\
    & 03:16:29.0 & 36.41 & 8.7\\
    & 03:16:32.5 & 35.89 & 8.6\\
    \hline
    \multirow{3}{*}{\textbf{ONEWEB-0203}} & 02:41:00.7 & 51.53 & 8.1\\
    & 02:43:01.2 & 83.22 & 7.6\\
    & 02:45:00.2 & 54.34 & 8.4\\
    \hline
    \multirow{3}{*}{\textbf{ONEWEB-0206}} & 02:58:59.2 & 46.22 & 8.4\\
    & 03:01:15.7 & 74.44 & 7.3\\
    & 03:03:27.9 & 49.80 & 8.2\\
    \hline
    \multirow{3}{*}{\textbf{ONEWEB-0207}} & 02:53:44.5 & 59.81 & 8.0\\
    & 02:55:15.0 & 81.98 & 7.6\\
    & 02:56:46.0 & 60.33 & 8.1\\
    \hline
    \multirow{8}{*}{\textbf{ONEWEB-0208}} & 02:17:56.7 & 29.26 & 9.4\\
    & 02:18:00.0 & 29.69 & 9.5\\
    & 02:18:03.5 & 30.13 & 9.3\\
    & 02:20:13.5 & 49.78 & 7.9\\
    & 02:21:45.6 & 59.08 & 7.6\\
    & 02:23:15.9 & 50.34 & 8.2\\
    & 02:24:29.4 & 38.55 & 6.8\\
    & 02:24:32.7 & 38.04 & 6.8\\
    \hline
    \multirow{2}{*}{\textbf{ONEWEB-0209}} & 02:29:00.4 & 50.45 & 8.3\\
    & 02:33:02.0 & 48.45 & 8.4\\
    \hline
    \end{tabular}
\end{table}

\begin{table}[htb]
    \centering
    \caption{Observational campaign: November 6$^{\rm{th}}$, 2021.}
    \label{tab: C4 - Nov}
    \begin{tabular}{|c|c|c|c|}
    \hline
    \textbf{Satellite} & \textbf{UTC} & \makecell{\textbf{Elevation}\\\textbf{[deg]}} & \makecell{\textbf{Apparent}\\\textbf{Magnitude}}\\
    \hline
    \textbf{ONEWEB-0102} & 17:58:58.8 & 77.75 & 8.4\\
    \hline
    \textbf{ONEWEB-0109} & 18:15:38.8 & 57.30 & 8.1\\
    \hline
    \textbf{ONEWEB-0110} & 17:00:01.8 & 40.91 & 8.0\\
    \hline
    \textbf{ONEWEB-0111} & 17:42:25.2 & 84.09 & 8.0\\
    \hline
    \textbf{ONEWEB-0114} & 18:29:04.4 & 45.78 & 8.4\\
    \hline
    \textbf{ONEWEB-0117} & 18:19:01.5 & 54.41 & 8.3\\
    \hline
    \textbf{ONEWEB-0118} & 17:12:36.2 & 52.41 & 7.7\\
    \hline
    \textbf{ONEWEB-0121} & 18:12:24.1 & 61.10 & 8.0\\
    \hline
    \textbf{ONEWEB-0122} & 18:02:19.1 & 71.48 & 8.0\\
    \hline
    \textbf{ONEWEB-0123} & 17:55:46.1 & 79.60 & 8.1\\
    \hline
    \textbf{ONEWEB-0125} & 17:06:01.1 & 46.49 & 7.8\\
    \hline
    \textbf{ONEWEB-0128} & 17:19:15.8 & 58.25 & 7.6\\
    \hline
    \textbf{ONEWEB-0130} & 17:29:12.2 & 68.54 & 7.8\\
    \hline
    \textbf{ONEWEB-0131} & 17:25:50.9 & 65.16 & 7.7\\
    \hline
    \textbf{ONEWEB-0132} & 18:09:00.6 & 64.65 & 8.2\\
    \hline
    \textbf{ONEWEB-0134} & 17:45:46.7 & 88.37 & 8.0\\
    \hline
    \multirow{2}{*}{\textbf{ONEWEB-0135}} & 17:02:39.1 & 44.18 & 7.9\\
    & 17:02:42.6 & 44.18 & 7.9\\
    \hline
    \textbf{ONEWEB-0136} & 17:32:31.4 & 72.54 & 7.6\\
    \hline
    \textbf{ONEWEB-0137} & 18:25:45.1 & 48.96 & 8.3\\
    \hline
    \multirow{2}{*}{\textbf{ONEWEB-0139}} & 17:09:14.3 & 49.04 & 7.6\\
    & 17:09:17.6 & 49.04 & 7.6\\
    \hline
    \textbf{ONEWEB-0140} & 17:39:07.2 & 80.02 & 7.8\\
    \hline
    \textbf{ONEWEB-0141} & 17:27:31.9 & 66.63 & 7.7\\
    \hline
    \textbf{ONEWEB-0144} & 18:05:40.4 & 67.52 & 8.2\\
    \hline
    \multirow{2}{*}{\textbf{ONEWEB-0145}} & 17:22:29.0 & 61.44 & 7.8\\
    & 17:22:32.5 & 61.44 & 7.8\\
    \hline
    \textbf{ONEWEB-0146} & 17:15:56.5 & 54.94 & 7.8\\
    \hline
    \textbf{ONEWEB-0147} & 18:22:21.3 & 51.69 & 8.2\\
    \hline
    \end{tabular}
\end{table}

\end{appendix}

\end{document}